\documentclass[12pt, twocolappendix, numberedappendix]{openjournal}
\usepackage{blindtext}
\usepackage{mathtools}
\usepackage{xcolor}
\usepackage{textgreek}
\usepackage[utf8]{inputenc}
\usepackage[english]{babel}
\usepackage{hyperref}
\hypersetup{
    colorlinks=true,
    linkcolor=blue,
    filecolor=blue,      
    urlcolor=red,
    citecolor=blue,
}
\usepackage{color,colortbl}
\usepackage{tensind}
\tensordelimiter{?}

\DeclareGraphicsExtensions{.bmp,.png,.jpg,.pdf}
\usepackage{courier}
\usepackage{verbatim}
\usepackage{natbib}
\usepackage{tabularx}
\usepackage[normalem]{ulem}
\usepackage{orcidlink}
\usepackage{soul}

\graphicspath{ {./figs/} }

\begin{document}
\title{Bayesian analysis of a Unified Dark Matter model with transition: can it alleviate the \texorpdfstring{$H_{0}$}{Lg} tension?}

\author{Emmanuel Frion\orcidlink{0000-0003-1280-0315}}
\email{efrion@uwo.ca}
\affiliation{Helsinki Institute of Physics, P.O. Box 64, FIN-00014 University of Helsinki, Finland}
\affiliation{Department of Physics and Astronomy, Western University,
N6A 3K7, London, Ontario, Canada}
\author{David Camarena}
\affiliation{Department of Physics and Astronomy, University of New Mexico, Albuquerque, New Mexico 87106, USA}
\author{Leonardo Giani}
\affiliation{The University of Queensland,
School of Mathematics and Physics, QLD 4072, Australia}
\author{Tays Miranda}
\affiliation{Helsinki Institute of Physics, P.O. Box 64, FIN-00014 University of Helsinki, Finland}
\affiliation{Department of Physics, P.O.Box 35 (YFL), FIN-40014 University of Jyv$\ddot{a}$skyl$\ddot{a}$, Finland}
\author{Daniele Bertacca\orcidlink{0000-0002-2490-7139}}
\affiliation{Dipartimento di Fisica e Astronomia “G. Galilei”, Università degli Studi di Padova, via Marzolo 8, I-35131, Padova, Italy}
\affiliation{INFN, Sezione di Padova, via Marzolo 8, I-35131, Padova, Italy}
\affiliation{INAF - Osservatorio Astronomico di Padova, Vicolo dell’Osservatorio 5, I-35122 Padova, Italy}
\author{Valerio Marra\orcidlink{0000-0002-7773-1579}}
\affiliation{Núcleo de Astrofísica e Cosmologia \& Departamento de Física, Universidade Federal do Espírito Santo, 29075-910, Vitória, ES, Brazil}
\affiliation{INAF -- Osservatorio Astronomico di Trieste, via Tiepolo 11, 34131 Trieste, Italy}
\affiliation{IFPU -- Institute for Fundamental Physics of the Universe, via Beirut 2, 34151, Trieste, Italy}
\author{Oliver F. Piattella}
\affiliation{Dipartimento di Scienza e Alta Tecnologia, Università degli Studi dell’Insubria e INFN, via Valleggio 11, I-22100 Como, Italy}

\begin{abstract}
We consider cosmological models in which Dark Matter (DM) and Dark Energy (DE) are described by a single component, dubbed Unified Dark Matter (UDM) models, in which the DE-like part can have an equation state $<-1$ at late times without violating the null energy condition. In this paper, we investigate whether this feature can relieve the Hubble tension. We perform a Bayesian analysis of the model using SNIa data from Pantheon, the CMB distance prior from Planck, and the prior on the absolute magnitude $M$ of SNIa from SH0ES. Using the prior, the data suggests a smooth transition taking place at redshifts $z_{\rm t} \simeq 2.85$, which provides a value $H_0=69.64\pm 0.88$ for the Hubble constant, slightly alleviating the tension by $\sim 1.5 \sigma$. Without it, we obtain $H_0 = 67.6^{+1.3}_{-0.82}$ and a transition happening at $z_t=1.36$. We also discuss the importance of using the prior on $M$ for constraining this model.
\end{abstract}

\maketitle

\onecolumngrid
\twocolumngrid
\onecolumngrid
\twocolumngrid

\section{Introduction}
\label{sec:int}

\twocolumngrid

The observed accelerated expansion of the Universe requires, within the framework of General Relativity, some form of Dark Energy (DE) to overcome the gravitational collapse of ordinary matter.
A cosmological constant $\Lambda$ seems to be the most natural candidate for DE, and together with Cold Dark Matter (CDM) they constitute the main ingredients of the standard model of cosmology, hereby referred to as $\Lambda$CDM.

Despite providing an extremely successful and (relatively) simple description of the expansion history of the Universe, the $\Lambda$CDM model has been recently challenged by the appearance of statistical tensions between the values of two cosmological parameters measured using late- and early-times probes. Specifically, there is a $\sim 5 \sigma $ tension concerning the value of the Hubble factor today $H_0$, and a 2--3$\sigma$ tension in the parameter combination $S_8\equiv \sigma_8\left(\Omega_{{\rm m}0}/0.3\right)^{0.5}$, where $\sigma_8$ is the averaged amplitude of the linear matter density fluctuations over spheres of radius  $8 h^{-1}$ Mpc today and $\Omega_{{\rm m}0}$ is  the present day matter density. Early-times probes seem to prefer lower values of $H_0$ and higher values of $S_8$ than late-times ones, see for example \cite{DiValentino:2021izs,Schoneberg:2021qvd,Abdalla:2022yfr,Perivolaropoulos:2021jda,Bernal:2016gxb,DiValentino:2020vvd} for a review of these problems.

It is worth noticing that the $H_0$ tension might be also interpreted as a tension on the absolute magnitude $M$ of type Ia supernovae, since the calibration of the absolute magnitude is inferred from the luminosity-distance relation of supernovae at both high and low redshift, therefore  introducing correlations between the value of $M$ and the intrinsic properties of DE \cite{Camarena:2019moy,Benevento:2020fev,Camarena:2021jlr,Efstathiou:2021ocp, Camarena:2023rsd}. If not due to systematics\footnote{In particular, the $H_0$ tension might be related to systematics in supernova standardization \cite{Wojtak:2023sts} or in the Cepheid calibration of the cosmic ladder. For example, the analysis of \cite{Freedman:2019jwv} uses Type Ia supernovae (SnIa) observations calibrated with the Tip of the Red Giant Branch \cite{TRGB1,TRGB2} rather than Cepheids, and results in $H_0 = 69.8 \pm 0.8\; \text{(stat)} \pm 1.8\; \text{(sys)}\; \text{km}\; \text{s}^{-1} \text{Mpc}^{-1}$, compatible with the value inferred by the Planck collaboration $H_0 = 67.4 \pm 0.5 \text{km}\; \text{s}^{-1} \text{Mpc}^{-1}$ \cite{Planck:2018vyg}. Note that systematics in Cepheids only are insufficient to solve the tension though \cite{Riess:2020xrj,Sharon:2023ioz}.} or to a possible redshift evolution of cosmological parameters \cite{Krishnan:2020vaf,Dainotti:2021pqg,Dainotti:2022bzg,Lenart:2022nip,Colgain:2022rxy,Malekjani:2023dky}, these observations will require new physics beyond $\Lambda$CDM to be properly addressed \cite{Hu:2023jqc}.  On the other hand, it is unclear which kind of new physics could successfully tackle both tensions \textit{at the same time} \cite{Anchordoqui:2021gji,Schoneberg:2021qvd}.
Indeed, naive resolutions of one seem to worsen the other. For example, if one tries to solve the $H_0$ tension at late times by increasing the present day DE energy density $\Omega_{{\rm DE}0}$, then the matter density decreases proportionally ($\Omega_{{\rm m}0} \approx 1 - \Omega_{{\rm DE}0}$ today), and consequently $S_8$ decreases, exacerbating the $S_8$ tension. 

Most of the attempts addressing the $H_0$ tension can be classified into early- and late-time modifications of the $\Lambda$CDM expansion history \cite{Schoneberg:2021qvd}. Early-time modifications aim to modify the value of the sound horizon $r_{\rm s}$ \cite{Murgia:2020ryi,Poulin:2018cxd,Poulin:2021bjr,Smith:2019ihp,Smith:2020rxx,Sabla:2021nfy,Sabla:2022xzj,Capparelli:2019rtn,Jedamzik:2020krr, Bamba:2012cp}, which  results in a different value of $H_0$ inferred from the CMB\footnote{Though an increased $H_0$ within this approach usually creates a tension with $\Omega_{\rm m 0} h^2$, with either BAO or weak lensing data \cite{Jedamzik:2020zmd}}. Late-time modifications instead try to obtain a higher $H_0$ by modifying the expansion history at recent times, for example including interactions in the dark sector, through dynamical DE models, or through dark matter models with varying equation of state \cite{Renk:2017rzu,DiValentino:2017iww,Yang:2018euj,Pan:2019gop,vonMarttens:2019ixw,Naidoo:2022rda,Montani:2023xpd}. Gravitational transitions models, in which the effective gravitational coupling $G$ undergoes a rapid transition at low redshift, have also been proposed as a resolution of the Hubble tension \cite{Alestas:2021nmi,Alestas:2022xxm,Marra:2021fvf,Zhou:2021xov} because they change the value inferred for the absolute magnitude of type Ia supernovae $M$, therefore providing a better fit than smooth $H(z)$ models \cite{Alestas:2021luu}. 
Concerning late time resolutions, the analysis of \cite{Heisenberg:2022gqk,Heisenberg:2022lob,Lee:2022cyh} indicates that, in order to not worsen the $S_8$ tension, a dynamical DE field is required with Equation of State (EoS) parameter evolving from  $w_{\rm DE} \geq-1$ to $w_{\rm DE} < -1$. 
Perfect fluids satisfying the second inequality are  labelled \textit{phantomic} since the seminal work \cite{Caldwell:1999ew}. They are considered  unphysical for multiple reasons. Among them, we mention that their kinetic energy is negative, therefore introducing instability at high energy, and also that their energy density grows with the expansion of the Universe, consequently undermining the principle of energy conservation \cite{Copeland:2006wr,Amendola:2015ksp}.

In this work, we consider a unified model for the dark sector, called Unified Dark Matter (UDM) model or {\it Quartessence}, of the Universe where DM and DE are interpreted as different manifestations of the same dark component. Many works have investigated this unification, which may appear, for instance, as a consequence of the Holographic principle \cite{Aviles:2011sfa} or via the addition of an adiabatic fluid \cite{Dunsby:2016lkw}. The potential of UDM models in addressing challenges to $\Lambda$CDM such as the cosmological constant problem \cite{DAgostino:2022fcx} makes this class of models particularly interesting. Here, we focus on a particular class of UDM models in which the DE-like part of the model can also mimic a {\it phantom} fluid behavior. UDM models were investigated extensively in the past, see for example \cite{Kamenshchik:2001cp,Bilic:2001cg,Carturan:2002si,Sandvik:2002jz,Gorini:2004by,Gorini:2007ta,Campos:2012ez,Piattella:2009da,Gorini:2009em,Bruni:2012sn} on the generalized Chaplygin gas, \cite{Bertacca:2007ux,Bertacca:2007cv,Bertacca:2007fc,Balbi:2007mz,Quercellini:2007ht,Bertacca:2008uf,Bilic:2008yr,2009camera,Li:2009mf,2010Gao,Lim:2010yk,2011camera,Piattella:2009kt,Bertacca:2010mt,Luongo:2018lgy, Boshkayev:2019qcx} concerning scalar field models, or more general, non-adiabatic models, see \textit{e.g.}\ \cite{Reis:2004hm}. More recent proposals were also given in \cite{Boshkayev:2021uvk,Mishra:2018tki,Leanizbarrutia:2017afj,Benisty:2018qed,Anagnostopoulos:2019myt}. The potential of UDM models in addressing the $S_8$ tension was investigated in \cite{Camera:2017tws}. 

Inspired by these models, in this work we consider the possibility of addressing the Hubble tension with a $w_{\rm DE}<-1$ of the DE-like part at late times, but evolving towards an asymptotic de Sitter attractor. The presence of the latter mitigates the stability issues by avoiding the appearance of a future big-rip singularity, see, for example, \cite{Singh:2003vx,Sami:2003xv,Capozziello:2005tf}, and is a key feature in many beyond $\Lambda$CDM scenarios, see for example \cite{Oriti:2021rvm,Hashim:2020sez,Roy:2017uvr,Capozziello:2005tf,Giani:2019xjf,Belgacem:2017cqo}. We restrict our study at the background level for this purpose, so we do not consider structure formation at this time. 

To illustrate how this type of models can potentially address the Hubble tension, we will employ a very simple toy model proposed originally in \cite{Bertacca:2010mt}, for which the UDM energy momentum can be described as a perfect fluid with a fixed, time-dependent, analytical pressure profile chosen \textit{a priori}. This can be done through a Lagrange multiplier, for example, or by fixing at the background level a suitable initial condition and a scalar Lagrangian field with a non-canonical kinetic term that can reproduce this pressure profile \cite{Bertacca:2008uf, Bertacca:2010ct, Bertacca:2010mt}.

The structure of the paper is the following: in Sec.~\ref{sec:udm}, we review the UDM model proposed in \cite{Bertacca:2010mt}, and  discuss under which conditions a simple toy model can address the Hubble tension. Then, we perform a statistical analysis of the chosen model in Sec.~\ref{sec:stat}. 
In Sec.~\ref{results}, we report our results and in Sec.~\ref{sec:concl} our conclusions.

\section{A simple UDM toy model}
\label{sec:udm}

Using the e-fold number $N = \log{a}$ as time parameter (here we set $a(t_0)=1$), the continuity equation of a UDM fluid in a FLRW background can be written as:
\begin{equation}\label{CeQ}
    \frac{d\rho(N)}{dN} +3\rho(N) = -3p(N)\;.
\end{equation}
Following \cite{Bertacca:2010ct}, for a given pressure $p(N)$, the formal solution of the latter equation is:
\begin{equation}\label{rhoUDM}
    \rho(N)= e^{-3N}\left[K  -3\int_{-\infty}^N d\bar{N} e^{3N}p(N)\right] \;,
\end{equation}
where $K$ is an integration constant. As a result, we see that UDM models always contain a dust-like component which corresponds to the homogeneous solution of Eq.~\eqref{CeQ}. The prescription above is very general and can be valid whether $p(N)$ is an analytic function or not. We immediately notice that $\Lambda$CDM, at the background level, can be a sub-case of this class of UDM models. Indeed if $p= {\rm const.}=-\Lambda$,  we are left with a Universe filled with a cosmological constant $\Lambda$ and  a dust fluid with $\rho(0)= K$ (see, \textit{e.g.}, \cite{Bertacca:2008uf}). Different choices of $p$ will result in  different behaviours, which make UDM models suitable to mimic a wide range of DE candidates.

Here, we adopt the ansatz proposed in~\cite{Bertacca:2010mt}, such that the pressure of the UDM fluid follows
\begin{equation}\label{pUDM}
p_\varphi= -\frac{\rho_\lambda}{2}\left\{1 + \tanh{\left[\frac{\beta}{3}\left(a^{3}-a_{\rm t}^3\right)\right]}\right\}\;,
\end{equation}
where $\rho_{\rm \lambda}$ is the energy density of an effective cosmological constant, such that $\rho_{\rm \lambda} \propto \Lambda$ with $\Lambda$ being the cosmological constant. After integration of Eq.~\eqref{CeQ}, this ansazt leads to the following density profile:
\begin{equation}\label{rhophi}
\rho_\varphi = \frac{\rho_\lambda}{2} +\frac{3}{a^3\beta}\frac{\rho_\lambda}{2}\ln{\left\{\cosh{\left[\frac{\beta}{3}\left(a^{3}-a_{\rm t}^3\right)\right]}\right\}} + \frac{\rho_{{\rm m}0}}{a^{3}}\,. 
\end{equation}
In order to understand the phenomenology of $\rho_\varphi$, we split the UDM fluid and isolate the contribution of the effective DE component, \textit{i.e.} the energy density that {\it a  posteriori} we interpret as a DE-like component at the background level, and of the DM one. Specifically, we have defined $\rho_{{\rm m}}(a)=\rho_{{\rm m}0}a^{-3}$ as the energy density of the effective DM-like component and, consequently $\rho_{\mathrm{\rm DE}} = \rho_\varphi - \rho_{{\rm m}0}a^{-3}$ as the energy density of the effective DE-like component. Bearing in mind that the matter component does not contribute to the pressure, we additionally define the EoS parameter of the DE-like component as $w_{\mathrm{\rm DE}} = p_\varphi/\rho_{\mathrm{\rm DE}}$.

\begin{figure}[!tbp]
    \centering
	\includegraphics[width=0.99\columnwidth]{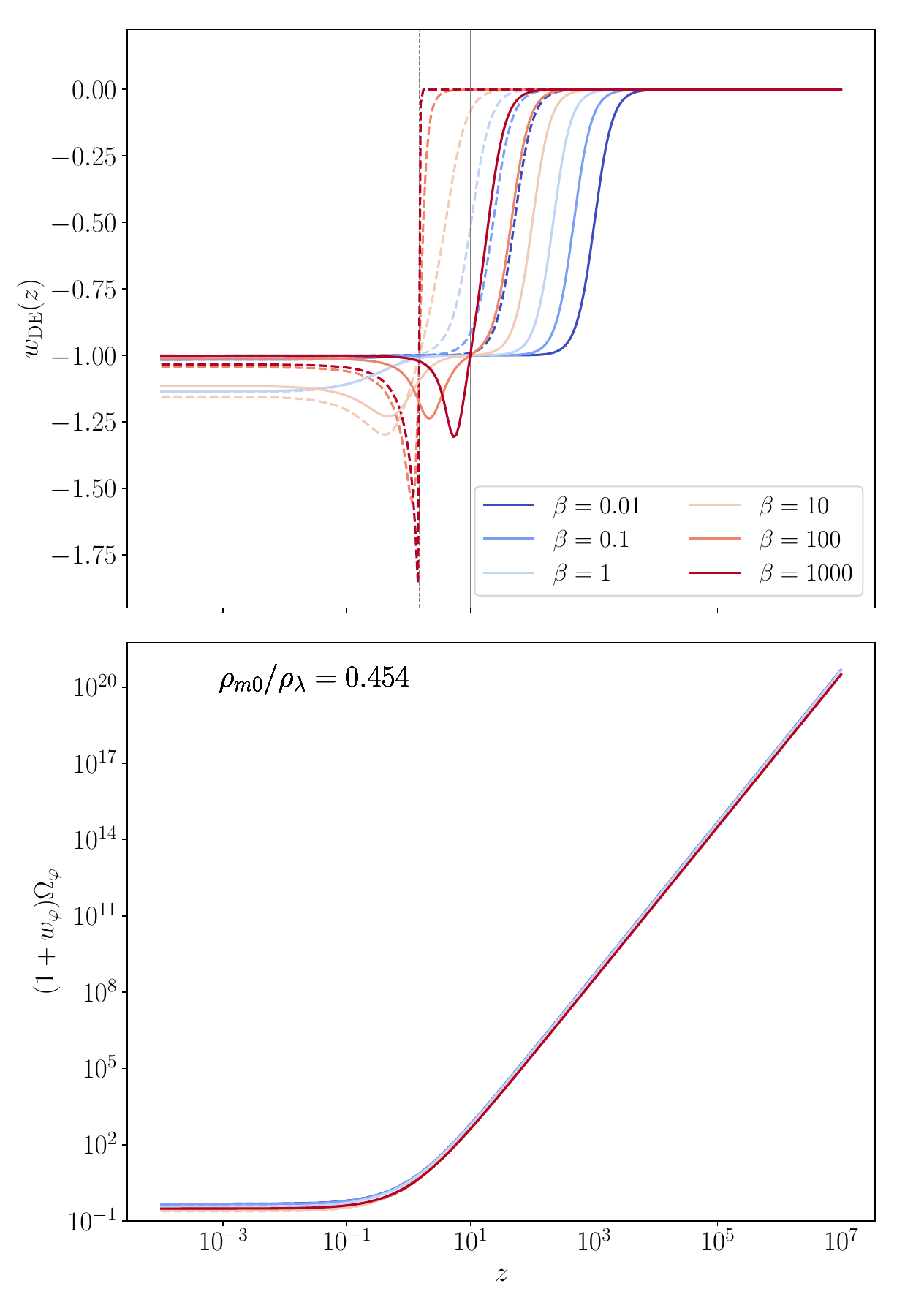}\\
	\caption{$w_{\mathrm{\rm DE}}$ (top panel) and the WEC condition (bottom panel) as functions of the redshift for different combinations of $\beta$ and $a_t$. Solid and dashed color lines represent the corresponding function evaluated at $a_t = 1/11$ (vertical black solid line) and  $a_t = 0.4$ (dashed black solid line), respectively. The lighter blue and red lines ($\beta = 1, 10$) show that the UDM fluid can introduce a phantomic dark energy component without violating the WEC. We maintain this color and line-style notation throughout the manuscript, unless otherwise specified.}
	\label{fig:wz_wec}
\end{figure}

With the above profile, this UDM model typically transitions from a matter fluid to a fluid dominated by a (phantomic) DE component. The moment at which the field starts behaving differently than either a pure matter component or a pure DE field occurs at time $a_t$, and the speed of the transition is controlled by $\beta$. Such behaviour is shown in the top panel of Fig.~\ref{fig:wz_wec} where we show $w_{\mathrm{\rm DE}}$ as a function of the redshift for different sets of values of $\beta$ and $a_{\rm t}$  (see also Figs. in \cite{Bertacca:2010mt}). At high redshifts, all curves merge to $w_{DE}=0$. Depending on the values of $\beta$ and $a_t$, $w_{\mathrm{\rm DE}}(z)$ transitions from a matter-like fluid to an effective dark energy component that can attain the phantomic regime. This is particularly noticeable for intermediate  values of $\beta$, \textit{i.e.}, $\beta \in [1,10]$. At low redshifts, $\textit{i.e.}$ in the limit $a\gg1$, the $w_{DE}$ curves remain constant at a different value. The constant behaviour from the high- and low-redshift regimes reflect the asymptotic limit of the hyperbolic tangent function we chose as ansatz. As we discuss later in this section, the smoothness and transition to the $\Lambda$CDM regime is controlled by the combination of $\beta$ and $a_{\rm t}$.

The division of the UDM fluid into the dark energy and matter components is justified from a phenomenological point of view, since an observer can associate the DM and DE components of this UDM model with the observed ones. Indeed,  since the hyperbolic cosine function is bounded by $1$ from below, the energy density $\rho_\mathrm{\rm DE}$ is a positive definite function that can be used to provide an effective interpretation of the dark energy sector. Consequently, $\rho_\varphi$ is also positive definite and, notably, the UDM fluid satisfies the Weak Energy Condition (WEC) as long as $\rho_\varphi + p_\varphi \geq 0$, or equivalently: 
\begin{gather}\label{WECphi}
(1+w_{\rm DE})\left[\frac{\rho_\lambda}{2} +\frac{3}{2\beta}\frac{\rho_\lambda}{a^3}\ln{F(\beta,a_t,a)} \right] + \frac{\rho_{\rm m0}}{a^3}\geq 0\,,
\end{gather}
where, for the sake of simplicity, we have defined
\begin{align}
    F(\beta,a_t,a) \equiv \left\{\cosh{\left[\frac{\beta}{3}\left(a^{3}-a_t^3\right)\right]}\right\}\,.
\end{align}
Therefore, the UDM fluid can provide a phantomic dark energy component, \textit{i.e.} $w_{\mathrm{\rm DE}} < -1$, while keeping the concordance with the WEC. The bottom panel of Fig.~\ref{fig:wz_wec} shows the WEC, recast as $(1+w_{\varphi})\Omega_\varphi \geq 0$ with $w_\varphi = p_\varphi/\rho_\varphi$ and $\Omega_\varphi =[\rho_{\rm m0}a^{-3}+ \rho_{\rm DE}(1+w_{\rm DE})]/\rho_{\rm crit}$ normalised with the critical density $\rho_{\rm crit}$ at $a=1$, as a function of redshift for different values of $a_{\rm t}$ and $\beta$. The lightest blue and red lines ($\beta = 1$ and $\beta = 10$) in Fig.~\ref{fig:wz_wec} illustrate that the UDM fluid can behave as a phantomic DE component without violating the WEC. 

In a flat FLRW background, the corresponding Friedmann equations (in units $8\pi G=1$) at late times, for which we neglect the radiation component, may be written as:
\begin{eqnarray}\label{Feq}
3H^2 &=& \frac{\rho_{\rm \lambda}}{2} + \rho_{\rm m} + \rho_{\rm comp}\; ,\\
3\frac{\ddot{a}}{a} &=& \frac{\rho_{\rm \lambda}}{4} - \frac{\rho_{\rm m}}{2} -\frac{1}{2}\left(\rho_{\rm comp} + 3p_{\rm comp}\right)\;\label{Acceq},
\end{eqnarray} where we have further split the effective dark energy component into a vacuum component $\rho_{\lambda}$ and the convenient complementary quantities $\rho_{\rm comp}$ and $p_{\rm comp}$ as 
\begin{eqnarray}\label{rhocomp}
\rho_{\rm comp} &:=& \frac{3}{a^3\beta}\frac{\rho_\lambda}{2}\ln{\left\{\cosh{\left[\frac{\beta}{3}\left(a^{3}-a_t^3\right)\right]}\right\}} \,,\\
p_{\rm comp}&:=& -\frac{\rho_{\rm \lambda}}{2}\tanh{\left[\frac{\beta}{3}\left(a^{3}-a_t^3\right)\right]} \,.
\label{pcomp}
\end{eqnarray}
These equations highlight the physical significance of the $a_t$ parameter. When $a<a_t$, the complementary component is a perfect fluid with positive pressure, whereas for $a >a_t$ its pressure is negative. Therefore, $a_t$ indicates the moment in the cosmological history where the EoS parameter of the complementary contribution, $w_{\rm{comp}}=p/\rho_{\rm{comp}}$, changes sign. Note that in the limiting case, $a=a_t$, the EoS parameter is well defined $w_{\rm{comp}}=0$, but the complementary component does not contribute to the field equations since both its pressure and density vanish.

Written in this form, the UDM fluid recovers the standard $\Lambda$CDM phenomenology today in two scenarios:\footnote{In this work, we consider the evolution of the Universe only until today, so that $0<a\leq1$. However, UDM models in which $a\rightarrow \infty$ can also recover the $\Lambda$CDM limit.}
\begin{enumerate}
     \item[(i)] when the contributions of the complementary component are negligible, \textit{i.e.} $\rho_{\rm comp} \rightarrow 0$ and $p_{\rm comp} \rightarrow 0$, and the effective cosmological constant is set to $\rho_\lambda = 2\Lambda$\,, or \\[-0.5cm]
     \item[(ii)] when they tend towards  $\rho_{\rm comp} \rightarrow \rho_\lambda/2$ and $p_{\rm comp} \rightarrow -\rho_\lambda/2$, with an effective cosmological constant $\rho_\lambda = \Lambda$\,.
\end{enumerate}

Since one of our goals is to address the potentiality of this model in tackling the Hubble tension without spoiling the observational success of the $\Lambda$CDM model, in what follows we expand the discussion on the aforementioned  $\Lambda$CDM limits. We explicitly demonstrate that such limits are achieved either by $| \beta (a^3-a_t^3)/3 | \ll 1$ or $
|\beta (a^3-a_t^3)/3| \gg 1$, where the former leads to $\rho_{\rm comp} \rightarrow 0$ and the latter to $\rho_{\rm comp} \rightarrow \rho_\Lambda/2$. We will show that, in these limits, the complementary component can be decomposed into components acting like either a cosmological constant, usual matter, or an exotic form of matter. This decomposition allows us to add the contributions of the complementary component like small deviations of the $\Lambda$CDM expansion. Even though we do not know precisely when this decomposition occurs, it gives us a convenient way to investigate the importance of the complementary term,  particularly at late times. In order to lighten the notation, hereafter we define $\alpha := \beta (a^3-a_t^3)/3$.

\begin{figure}[!tbp]
    \centering
    \includegraphics[width=0.99\columnwidth]{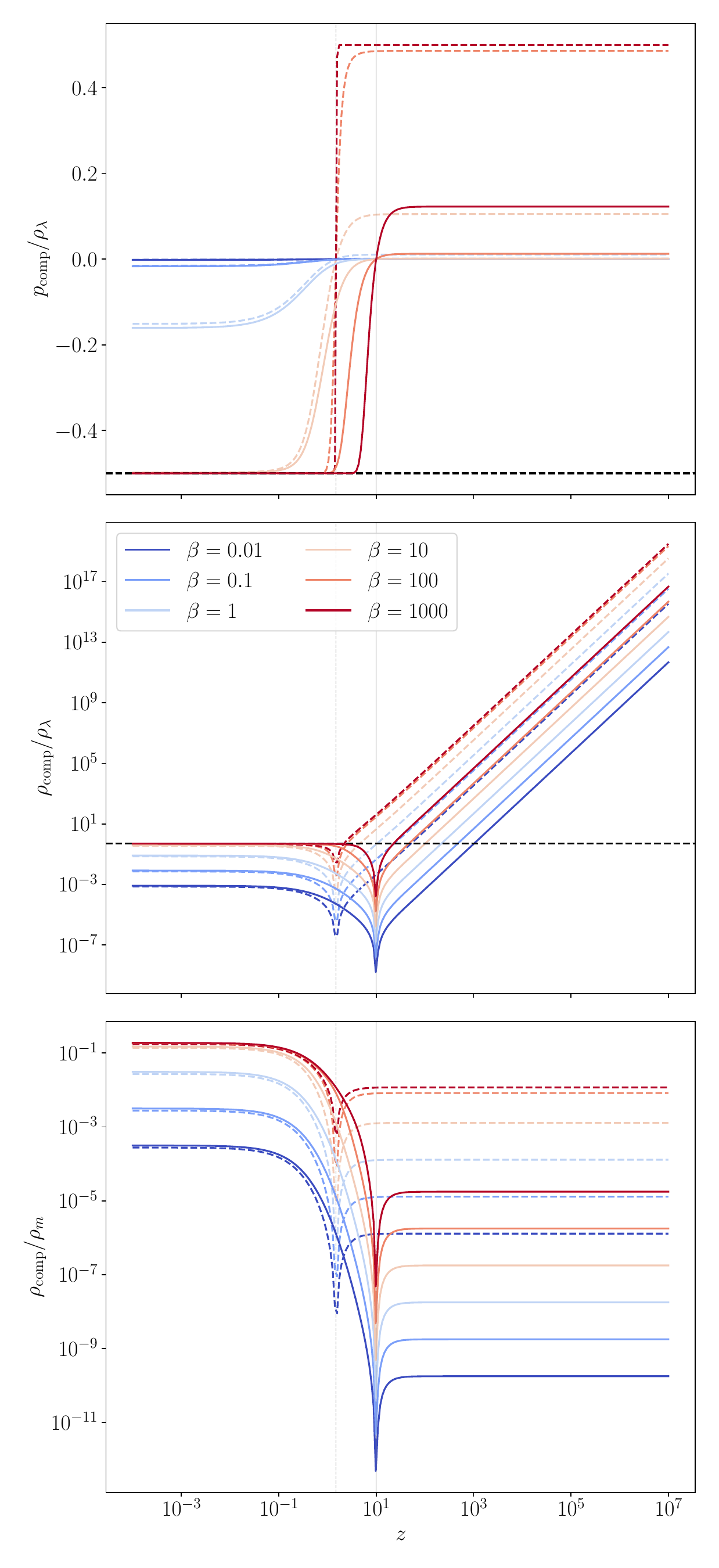}
    \caption{$p_{\rm comp}/\rho_\lambda$ (top panel), $\rho_{\rm comp}/\rho_\lambda$ (middle panel), and $\rho_{\rm comp}/\rho_{\rm m}$ (bottom panel) as functions of the redshift for different combinations of $\beta$ and $a_t$. Solid and dashed color lines represent the corresponding functions evaluated at $a_t = 1/11$ and  $a_t = 0.4$, respectively. The vertical solid and dashed black lines represent $z_t = 10$ and $z_t = 1.5$, respectively. On the horizontal dashed lines, we have $\rho_{comp}=\rho_{\lambda}/2$, for which the $\Lambda$CDM limit is recovered .}
    \label{fig:rho_comp}
\end{figure}

\subsection{Slow transition (\texorpdfstring{$ | \alpha| \ll 1$}{})}
Let us consider the Taylor expansion of Eqs.~\eqref{rhocomp} and \eqref{pcomp} for $|\alpha| \ll 1$. Their leading order contributions are 
 \begin{eqnarray}
 p_{\rm comp} &\simeq&\label{pbetasmall} - \frac{\alpha }{2}\rho_\lambda\;,\\
 \rho_{\rm comp}&\simeq& \frac{3\alpha^2}{4\beta}\frac{\rho_\lambda}{a^3} \,. 
\label{rhobetasmall}
\end{eqnarray}
These equations show that, in the slow transition limit, the contributions of the complementary component $p_{\rm comp}$ and $\rho_{\rm comp}$ are always small, and that the $\Lambda$CDM limit is reached if the effective cosmological constant is set to $\rho_\lambda = 2\Lambda$.

Even though the pressure of the complementary component is almost zero at all times, $\rho_{\rm comp}$ is not matter-like, as we can writing explicitly their expression for $\alpha$,
\begin{align}
    p_{\rm comp} &= -\frac{\beta}{6} (a^3-a_t^3) \rho_{\rm \lambda} \label{eq:pcompslowtrans}\;, \\
    \rho_{\rm comp} &= \frac{\beta}{3} \left( \frac{a^3}{2} - a_t^3 + \frac{a_t^6}{2a^3} \right) \frac{\rho_\lambda}{2} \,.\label{eq:slowtrans}
\end{align}
Eq~\eqref{eq:slowtrans} shows that, much before the transition ($a^3 \ll a_t^3$), the complementary component is dominated by the matter-like component $\propto \beta a_t^6 \rho_\lambda/a^3$. Much after the transition ($a^3 \gg a_t^3$), the term $\propto \beta a^3$ will prevail over the other contributions. Depending on the particular value of $a_t$, this regime could be not reached today though, since $a>1$ would be necessary. Therefore, in scenarios with $a_t$  close to unity, the energy density of the complementary component has a non-negligible $\Lambda$-like contribution $\propto \beta a_t^3$ at $a =1$.

Fig.~\ref{fig:rho_comp} shows the evolution of the complementary contributions in the form of $p_{\rm comp}/\rho_{\rm \lambda}$ (top panel), $\rho_{\rm comp}/\rho_{\rm \lambda}$ (middle panel) and $\rho_{\rm comp}/\rho_{\rm m}$ (bottom panel). Let us focus for now on the darkest solid and dashed blue lines ($\beta = 0.01$ and $\beta=0.1$), which display the slow transition regime. From the top panel, as anticipated, we can note that $p_{\rm comp}/\rho_\lambda$ is constantly small across all the history of the Universe. The middle panel shows that the quantity $\rho_{\rm comp}/\rho_\lambda$ becomes negligible in the slow transition regime at late times, of the order of $\alpha^2$. Additionally, the bottom panel confirms the matter-like behavior of the complementary component before the transition. As discussed before, the matter contribution at this epoch is proportional to $a_t^6$. This is confirmed by comparing the ratio of the dashed ($a_t = 1/11$, which corresponds to a transition at redshift $z_t=10$)  and solid (for which we choose a randomly lower transition redshift value $a_t = 0.4$)  blue lines ($\beta = 0.01$) in the bottom panel, where these values of $a_t$ lead to a $10^4$ discrepancy in $\rho_{\rm comp}/\rho_{\rm m}$.

\begin{figure}
    \centering    \includegraphics[width=0.99\columnwidth]{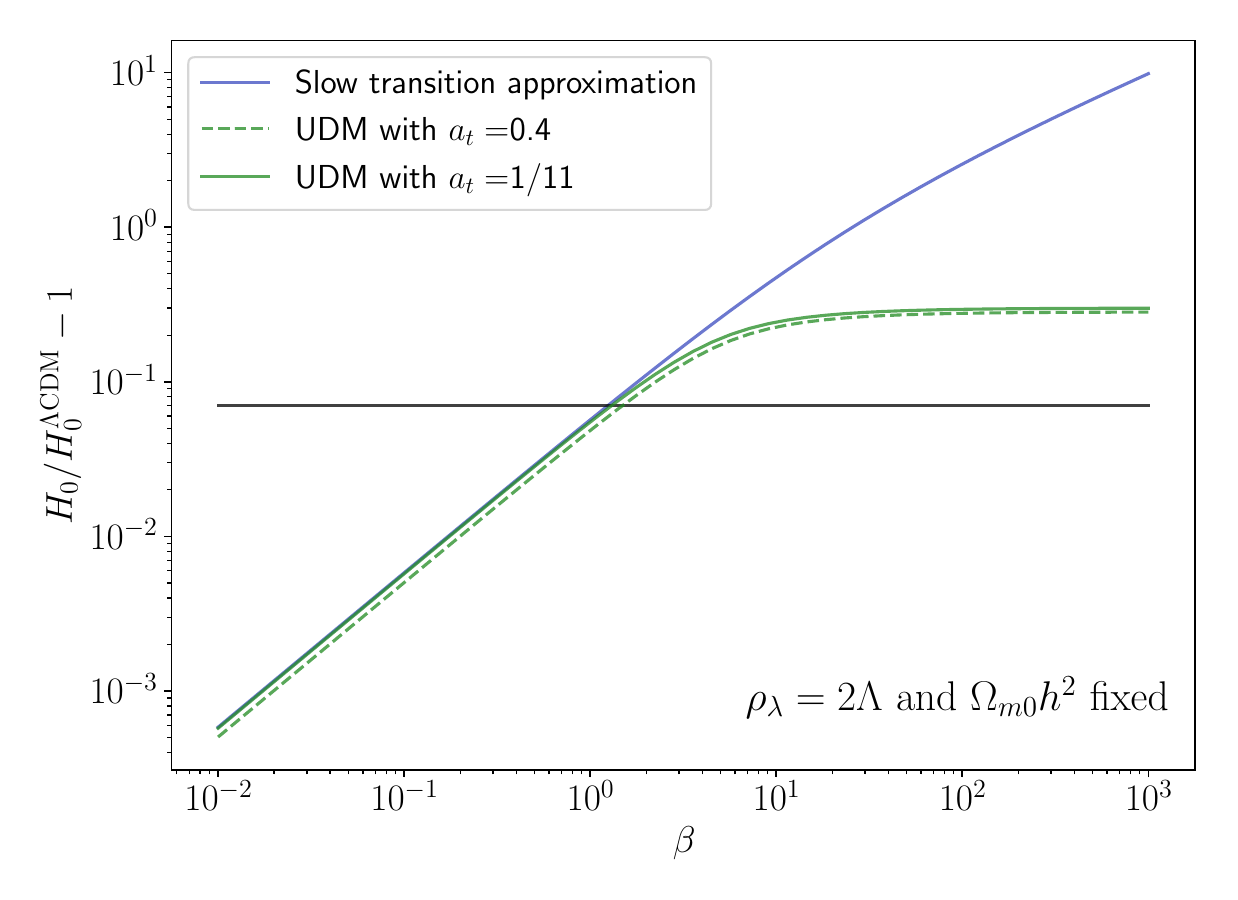}
    \caption{Relative change in the Hubble constant as a function of $\beta$ in the slow transition approximation (blue line) and from the numerical evaluation of the Friedmann equations (dashed and solid green lines). In the slow transition regime, the Hubble tension can be resolved with $\beta \sim 1$. Note that the horizontal black line represents $H_0=72$ km/s/Mpc, a value which alleviates the tension while keeping the expansion history close to $\Lambda$CDM at the percent level. In this plot, $H_0^{\Lambda CDM}=67.8$ km/s/Mpc.}
    \label{fig:deltaH_approx}
\end{figure}

Let us now address briefly which values of the parameters $\beta$ and $a_t$ are required to ease the Hubble tension in this scenario, \textit{i.e.}, for which values the complementary component could produce an expansion history very close to $\Lambda$CDM's own history, but with a percent level difference today. First, we rewrite the Friedmann equations in the slow transition regime by regrouping all contributions by powers of the scale factor, leading to
\begin{align}\label{smallbfeq}
3H^2 &\simeq \rho_{\rm \lambda,S} + \rho_{\rm m,S} +\rho_{\rm ph}\;, \\
3\frac{\ddot{a}}{a}&\simeq \rho_{\rm \lambda,S} -\frac{\rho_{\rm m,S}}{2} + \frac{5}{2}\rho_{\rm ph}\;, \label{smallbacceq}
\end{align}
where 
\begin{eqnarray}\label{smallbcc}
\rho_{\rm \lambda,S} &:=& \; \left(1-\frac{\beta}{3}a_t^3\right)\frac{\rho_\lambda}{2} \;,\\ \label{smallbmat}
\rho_{\rm m,S}&:=&\left(1 + \frac{\rho_\lambda}{\rho_{\rm m0}}\frac{\beta}{12}a_t^6\right)  \frac{\rho_{\rm m0}}{a^{3}} \;, \\ \label{smallbph}
\rho_{\rm ph}&:=&\frac{\beta}{12}a^{3} \rho_\lambda\;.
\end{eqnarray}
Equations \eqref{smallbcc}-\eqref{smallbph} show that, in the slow-transition regime, the cosmological constant-like $\rho_{\rm \lambda,S}$ and matter-like $\rho_{\rm m,S}$  terms  include complementary contributions and therefore deviate slightly from $\Lambda$CDM. The $\rho_{\rm ph}$ term includes contributions which act like neither a cosmological constant nor usual matter. It sources the phantomic contribution of the complementary fluid, hence the ``ph'' subscript, and will be used later in the paper to assess the capability of the UDM model to tackle the Hubble tension. From the expansion of the complementary pressure Eq.\eqref{eq:pcompslowtrans}, we define the phantomic pressure $p_{\rm ph}\simeq -\beta \rho_{\rm \lambda} a^3/6$, which includes the only exotic term in the total pressure (\textit{i.e.} the term which grows $\propto a^3$), and from which we derive the phantomic equation of state  $w_{\rm ph}:= p_{\rm ph}/\rho_{\rm ph}=-2$.  Although this phantomic field may dominate $\rho_{\rm comp}$ after the transition due to the small value of $\alpha$, the EoS parameter of the effective dark energy component remains consistent with $w_{\mathrm{\rm DE}}= -1$.
This is exemplified by the solid and dashed darker blues lines in the top panel of Fig.~\ref{fig:wz_wec}. Note that this phantomic component  appears only in the slow transition approximation of our UDM model, so that one implies the other.

\begin{figure*}
    \centering
    \includegraphics[width=\textwidth]{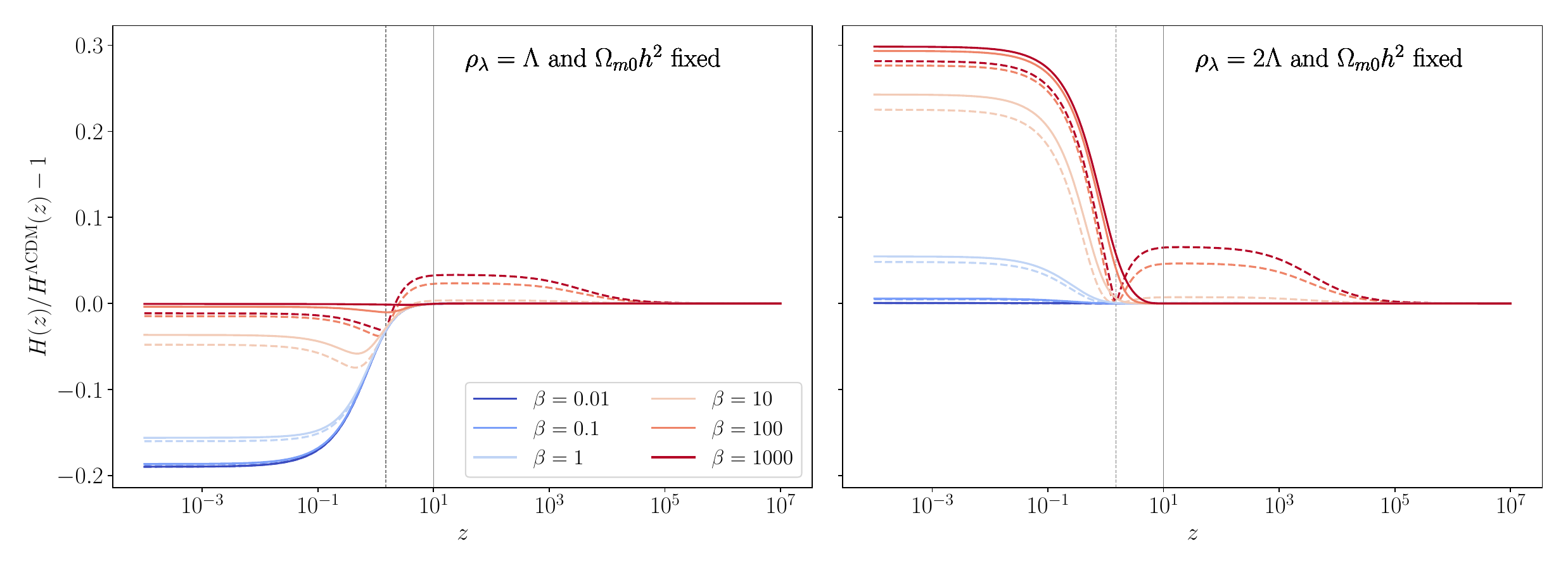}
    \caption{Relative variation of the Hubble function produced by the contribution of $\rho_{\rm comp}$ for two different values of the effective cosmological constant: $\rho_\lambda = \Lambda$ (left panel) and $\rho_\Lambda = 2 \Lambda$ (right panel). The darkest dashed and solid red lines ($\beta = 1000$ and $\beta = 100$) in the left panel show that the fast transition regime can, at most, recover the $\Lambda$CDM phenomenology for sufficiently small $a_t$. On the other hand, the darkest dashed and solid blue lines ($\beta = 0.01$ and $\beta = 0.1$) in the right panel represent the slow transition regime, where $\rho_{\rm comp}$ is of order $\alpha$. The lightest dashed and solid lines ($\beta = 1$) are solutions alleviating the tension while keeping an overall expansion rate consistent with the $\Lambda$CDM dynamics. Both panels show that large values of $a_t$ (dashed lines) significantly change $H(z)$ before the transition.}
    \label{fig:DH_rho_Lambdas}
\end{figure*}

Next, we choose the energy density of the effective dark energy component such that the $\Lambda$CDM limit is recovered for $|\alpha|\ll 1$ , \textit{i.e.}, we set $\rho_\lambda = 2 \Lambda$. Let us assume a sufficiently early transition, \textit{e.g.} $a_t \approx 1/11$, to avoid modifying the matter energy density at early times with a consequent impact on structure formation. From Eqs.~\eqref{smallbcc} and \eqref{smallbmat}, we conclude that the cosmological constant and  matter density shifts from the $\Lambda$CDM values proportionally to $a_t^3$ and $a_t^6$, \textit{i.e.}, by only about $10^{-3}\beta$ and $10^{-6}\beta$, respectively. The leading variation from the $\Lambda$CDM Hubble function is thus due to the remaining phantomic term 
\begin{equation}
    \rho_{\rm ph} (a=1) = \frac{\beta}{6}\Lambda \,,
\end{equation}
where we used the assumption $\rho_\lambda = 2 \Lambda$. Therefore, in the slow transition regime, the first Friedmann equation at $a=1$ becomes 
\begin{equation}
    3H_0^2 = 3 (H^{\Lambda \textup{CDM}}_0)^2 +\rho_{\rm ph}\,, 
\end{equation}
where $H^{\Lambda \textup{CDM}}_0$ is the Hubble constant in the $\Lambda$CDM limit such that  $3(H^{\Lambda \textup{CDM}}_0)^2 = \rho_{\rm m0}+\Lambda$. The shift in the Hubble constant, $\delta H_0 := H_0-H^{\Lambda \textup{CDM}}_0$, is given by
\begin{equation} \label{eq:deltaH_slow}
    \frac{\delta H_0}{H^{\Lambda \textup{CDM}}_0} = \left( \sqrt{1 + \frac{\beta}{6} \frac{\Lambda}{3(H^{\Lambda \textup{CDM}}_0)^2} } - 1 \right) \,.
\end{equation}
We compare in Fig.~\ref{fig:deltaH_approx} the shifts predicted by the slow transition regime as a function of $\beta$ (blue line) and the numerical resolution of Friedmann equation Eq.~\eqref{Feq} (green lines). Overall, Eq~\eqref{eq:deltaH_slow} offers a good approximation for values of $\beta \ll 1$ and $a_t$ sufficiently small, although it tends to slightly overestimate the cases with greater $a_t$, see for instance the dashed green line in Fig.~\ref{fig:deltaH_approx}.
Notably, the slow transition approximation holds even for values of $\beta \approx 1$. This is explained by the fact that the next-to-leading order term in Eq.~\eqref{rhobetasmall} scales as $\alpha^4$, which is also negligible for such $\beta$. Additionally, the shifts in the cosmological constant and matter energy densities are negligible too in this case.

Finally, we note from Fig.~\ref{fig:deltaH_approx} that the slow transition regime (or small deviations of it, \textit{i.e.}, $\beta \gtrsim 1$) could alleviate the tension. The horizontal black line corresponds to $H_0 \simeq 72$, a lower threshold needed to explain the mean $H_0$ measurements from various supernovae type 1a would at 1$\sigma$ (see \textit{e.g.} figure 10 of \cite{Perivolaropoulos:2021jda})\footnote{Note that we focus on supernovae data in the statistical analyses of the next section.}. Interestingly, we see in the top panel of Fig.~\ref{fig:wz_wec} that the limit $\beta = 1$ (lightest blue line) features a phantomic dark energy component $w_{\mathrm{\rm DE}} < -1$. This is not surprising though since, after the transition, $\rho_{\rm comp}$ is dominated by the phantomic contribution $\rho_{\rm ph}$ with an EoS parameter $w_{\rm ph} = -2$. We remind that this phantomic behaviour does not violate
the WEC, as seen earlier.

\subsection{Fast transition  (\texorpdfstring{$|\alpha| \gg 1$}{})}

In the limit $|\alpha| \gg 1$, we can approximate  $\ln{\cosh{\alpha}} \simeq |\alpha| -\ln{2}+e^{-2|\alpha|}$ and $\tanh{\alpha} \simeq \mathrm{sgn}(\alpha) (1-2e^{-2|\alpha|})$, so we can rewrite the complementary density
\eqref{rhocomp} and pressure \eqref{pcomp} as
\begin{eqnarray}
\rho_{\rm comp} &=& \frac{\rho_\lambda}{2}\left\lbrace \left|1-\frac{a_t^3}{a^3}\right| - \frac{3}{a^3\beta}\left(\log{2} - e^{-2|\alpha|}\right)  \right\rbrace\;, \label{rhobetalarge}\\
p_{\rm comp} &=& -\mathrm{sgn}(\alpha) \frac{\rho_\lambda}{2} \left(1- 2e^{-2|\alpha|} \right)\;, \label{pbetalarge}
\end{eqnarray}
where, for the sake of  illustration, we have kept the sub-dominant terms in $\alpha$. From this set of equations, and given that $\beta$ is assumed to be positive, we note that the behaviour of the UDM model depends on whether $\alpha$ is positive or negative, \textit{i.e.}, is different after and before the transition. We analyse these two limits in the remainder of this section.

\subsubsection{Before the transition: \texorpdfstring{$a < a_t$}{}}
\label{subsec:B1}

Before the transition, the complementary fluid exhibits a positive pressure of $p_{\rm comp} \approx \rho_\lambda/2$. This term cancels out the expected contribution of the $\Lambda$-like component in the total pressure of the UDM field leading to $p_\varphi \approx 0$. Therefore, the UDM field is dominated by the matter component before the transition in the fast transition regime, and $\rho_{\rm m}$ is the sole component behind the expansion history. This is particularly noticeable if we rewrite the Friedmann equations as
\begin{eqnarray}
3H^2 &=& \rho_{\rm m,F} + \rho_{\rm nph}\;,\\
3\frac{\ddot{a}}{a} &=& -\frac{\rho_{\rm m,F}}{2} -\frac{\rho_{\rm nph}}{2}\left(1 - 2\beta a^3\right)\;,
\end{eqnarray}
where we have defined
\begin{eqnarray}
\rho_{\rm m,F} &:=& \frac{\rho_{\rm m0}}{a^3} + \left(a_{\rm t}^3 -\frac{3\log2}{\beta}\right)\frac{\rho_\lambda}{2a^3} \;,\\
\rho_{\rm nph}&:=&\frac{3\rho_\lambda e^{2\alpha}}{2\beta a^3}\;,
\end{eqnarray}
where we use the subscript ``nph'' to exacerbate that this exotic contribution will not behave as a phantomic component, opposed to exotic contribution in the slow transition regime.
From the definition of $\rho_{\rm m,F}$, it is clear that changes in the total energy budget will be dominated by the shift $\propto \rho_\lambda a_t^3/a^3 $ in the matter density, since the non-matter contribution $\rho_{\rm nph}$ is negligible. In summary,  a sufficiently large value of $a_t$ changes significantly $\rho_{\rm m}$ while introducing a non-phantomic component with negligible energy density.

\subsubsection{After the transition: \texorpdfstring{$a > a_t$}{}}
\label{subsec:B2}

After the transition, the complementary pressure $p_{\rm comp} \approx -\rho_\lambda/2$ behaves like a cosmological constant, accounting for half of the total pressure since $p_\varphi \approx -\rho_\lambda$.
The Friedmann equations then become
\begin{eqnarray}
3H^2 &=& \rho_\lambda + \rho_{\rm m,F} +\rho_{\rm nph}\;,\\
3\frac{\ddot{a}}{a} &=& \rho_\lambda - \frac{\rho_{\rm m,F}}{2} -\frac{\rho_{\rm nph}}{2}\left(1 +2\beta a^3\right)\;,
\end{eqnarray}
where we now define
\begin{eqnarray}
\rho_{\rm m,F} &:=& \rho_{\rm m} - \left(a_t^3 +\frac{1}{\beta}3\log{2} \right)\frac{\rho_\lambda}{2a^3}\;, \\
\rho_{\rm nph}&:=& \frac{3\rho_\lambda}{2\beta a^3}e^{-2\alpha}\;.
\end{eqnarray}
Similarly to the previous case, the leading order contributions of the complementary fluid in the fast transition regime are in the matter term, with an amplitude proportional to $a_t^3/a^3$.

Our analysis shows that, prior to the transition, there is no cosmological constant-like contribution from this UDM model. Instead, the UDM field acts like a dust-like component plus an additional complementary component, whose density decays exponentially with $\alpha$. On the other hand, after the transition, it behaves as a cosmological constant plus a very similar complementary component. In both cases, the complementary fluid is non-barotropic, with an EoS parameter $w_{\rm nph}= \text{sign}(\alpha)2\beta a^3/3$ and a negligible energy density. 

After the transition, we recover the $\Lambda$CDM paradigm whenever $\beta \gg 1$, $a_t^3 \ll 1$, and $\rho_\lambda = \Lambda$.  The shift in the energy budget of matter and the contribution of the non-barotopic fluid both become negligible across all the cosmological history. Unlike the slow transition regime, the fast transition regime can differ significantly from $\Lambda$CDM. For instance, if we assume $a_t =0.4$ and $\beta = 1000$, we satisfy the fast transition condition and attain the $\Lambda$CDM limit today, though at the cost of a large shift in the matter component that can be seen from the darkest dashed red lines in the bottom panel of Fig.~\ref{fig:rho_comp}, for which $\rho_{\rm comp} \simeq 10^{-2} \rho_{\rm m}$. This result is dependent on the transition time, with later transitions resulting in greater contributions from the complementary fluid.  If the transition happens earlier, for instance at $a_t = 1/11$, $\rho_{\rm comp}\leq 10^{-5} \rho_{\rm m}$ at high redshifts, leaving the energy density of matter pragmatically unaltered (darkest solid red lines in the bottom panel of Fig.~\ref{fig:rho_comp}). The darkest dashed and solid red lines in the top and middle panels of Fig.~\ref{fig:rho_comp} show that, after the transition, the $\Lambda$CDM limit is recovered in the fast transition regime, \textit{i.e.} $\rho_{\rm comp}/\rho_\lambda = -p_{\rm comp}/\rho_\lambda = 1/2$, regardless of the value of $a_t$. In conclusion, the fast transition regime can resolve the Hubble tension only at the cost of greatly changing the expansion history, and only recovers the $\Lambda$CDM phenomenology for small $a_t$. It therefore fails at resolving the Hubble tension. The failure of the fast transition regime is not surprising according to previous results about fast transitioning models \cite{Heisenberg:2022gqk,Heisenberg:2022lob}, which show that a phantomic behaviour is needed in order to increase the current value of the Hubble constant, and that fields with non-phantomic component are not suited to handle the Hubble tension.

For the sake of completeness, we plot in Fig.~\ref{fig:DH_rho_Lambdas} the Hubble function produced by the UDM fluid when compared to the $\Lambda$CDM case under the assumption of $\rho_\lambda = \Lambda$ (left panel) and $\rho_\lambda = 2\Lambda$ (right panel). In the fast transition regime, \textit{i.e.}, $\beta \gg 1$, and assuming $\rho_\lambda = \Lambda$, we recover the $\Lambda$CDM behaviour nearly across all redshifts if $a_t$ is sufficiently small (solid darker red line). A later transition leads to a significant change in the Hubble rate, due to the change in the matter component before transition (solid darkest red line). On the other hand, in the slow transition limit, \textit{i.e.}, $\beta \lesssim 1$, and assuming $\rho_\lambda = 2\Lambda$, the cosmic expansion is similar to the $\Lambda$CDM expansion with only a small deviation at late times, see darkest blue lines in the right panel. However, for values of $\beta$ outside this regime, the expansion of the Universe is greatly modified at high redshifts. 

\begin{figure}[tbp]
    \centering
    \includegraphics[width=0.92\columnwidth]{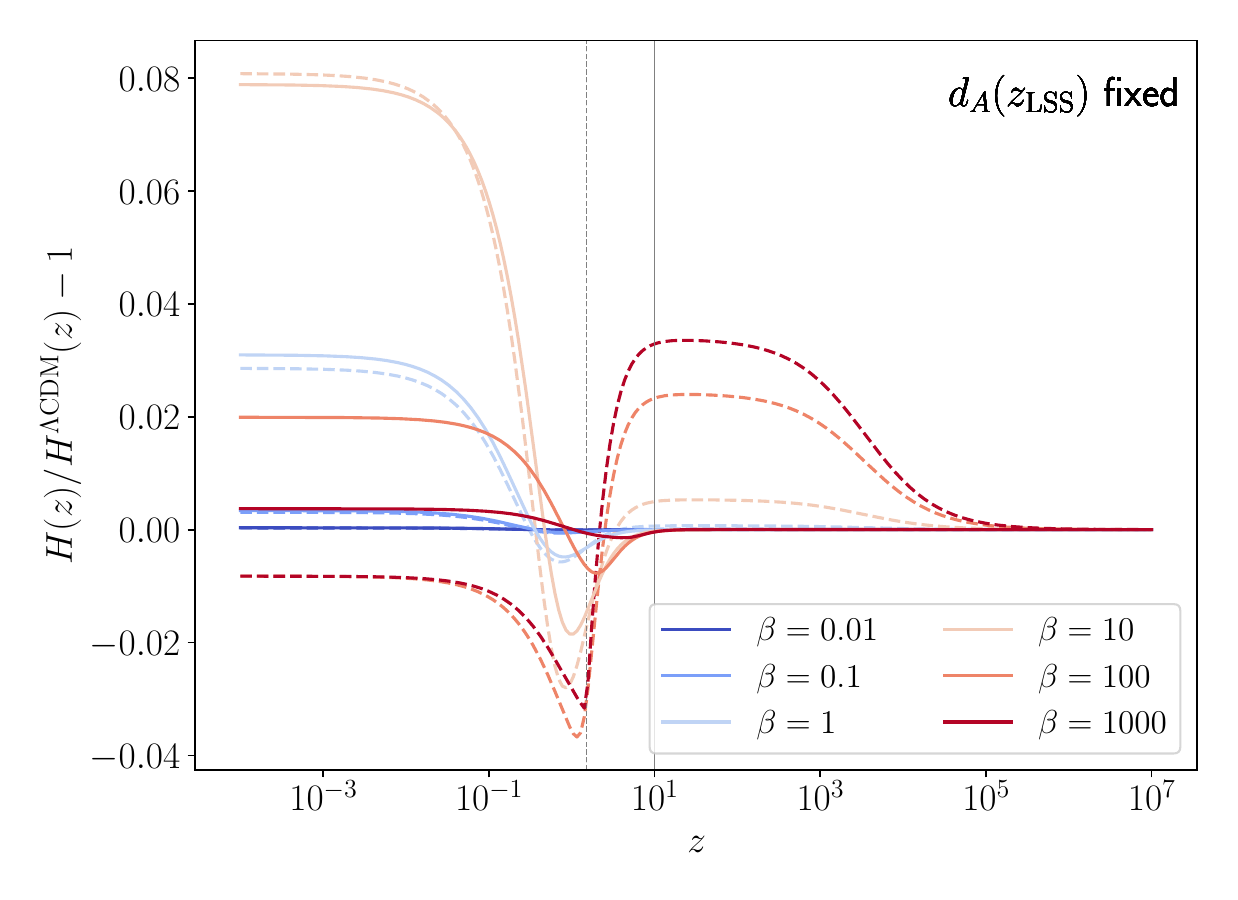}
    \caption{Relative variation of the Hubble function for different values of $\beta$ and $a_t$ when the angular distance to the last scattering surface is fixed to match the $\Lambda$CDM scenario.}
    \label{fig:DH_dA_fix}
\end{figure}

Finally, we insist on the fact that the chosen values of $\rho_\lambda$ in  Fig.~\ref{fig:DH_rho_Lambdas} are merely illustrations of the slow and fast transition regimes, and do not necessarily represent a realistic picture of the Universe. In order to present a more realistic scenario, we fix the values of $\rho_\Lambda$ and $H_0$ in  Fig.~\ref{fig:DH_dA_fix}, such that the angular distance to the last-scattering surface $D_{\rm A}(z_{\mathrm{LSS}})$ remains consistent with the value inferred from the $\Lambda$CDM constraints. We fix the time of last scattering at $z_{\mathrm{LSS}} = 1100$. In agreement with the deviation of the Hubble rate in the slow transition approximation \eqref{eq:deltaH_slow}, Fig.~\ref{fig:DH_dA_fix} shows that the Hubble tension is alleviated for intermediate values of $\beta \in [1,10]$, while presumably keeping a good agreement with the CMB observations.

\section{Statistical analysis}
\label{sec:stat}

Following the results of the previous section, we restrict our analysis to a parameter space  potentially solving the Hubble tension by imposing the flat priors  $\beta \in [0,10]$ and $a_t \in [0,1]$. Although these ranges are chiefly justified from the point of view of the Hubble tension, they prevent large modifications of the matter component, see Figs.~\ref{fig:DH_rho_Lambdas}~and~\ref{fig:DH_dA_fix}. This is crucial once we notice that large deviations in the matter field $\rho_{\rm m}$ at early times significantly modify the evolution of cosmological perturbations. Since we do not address the evolution of the cosmological perturbations of the UDM field, we effectively treat the UDM model as a late-time modification of the $\Lambda$CDM model.

We now use these assumptions to perform a Bayesian analysis of the UDM model considering cosmological probes of the background. Specifically, we consider SNIa data from the Pantheon catalog \cite{Scolnic:2017caz}, a prior on their absolute magnitude $M$ from Cepheids \cite{Camarena:2021jlr} and the CMB distance prior inferred from Planck data \cite{Chen:2018dbv}.

\subsection{CMB distance priors}
At the background level, the positions of the CMB acoustic peaks constrain cosmological distances through the so-called CMB distance prior. Typically, such prior is implemented via the baryon energy density $\Omega_{\rm b 0} h^2$, the spectral index $n_{\rm s}$, acoustic scale $l_{\rm A}$, and shift parameter~$R$:
\begin{align}
    l_{\rm A} &:= (1+z_{\star}) \frac{\pi D_{\rm A}(z_{\star})}{r_{\rm s}(z_{\star})} \;, \\
    R(z_{\star}) &:= (1+z_{\star}) \frac{D_{\rm A}(z_{\star}) \sqrt{\Omega_{\rm m 0} H_0^2}}{c} \;, \label{shift_factor}
\end{align}
where $z_{\star}$ is the decoupling redshift, $D_{\rm A}$ is the angular diameter distance, and $r_{\rm s}$ is the sound horizon\footnote{In the WMAP paper by Komatsu et al \cite{WMAP:2008lyn}, the authors argue that dark energy influences the distance scales and the growth of structures, though the sensitivity of the latter is limited. We show in Apendix \ref{ap:consistency} that the complementary fluid does not influence the CMB distance prior.}. Here, we assume a flat FLRW background, therefore $D_{\rm A}$ is
\begin{align}
    D_{\rm A}(z) = \frac{c}{(1+z) H_0} \int_{0}^{z} \mathrm{d}z^{\prime}  \frac{1}{E(z^{\prime})} \;,
\end{align}
where $E(z) \equiv H(z)/H_0$ is the normalised Hubble rate. As mentioned before, we adopt the CMB distance prior inferred from Planck 2018. Specifically, we use the values and correlation matrix presented in Table I of \cite{Chen:2018dbv} ($w$CDM model).\footnote{Note that although we do not deal with the matter power spectrum, the spectral index, $n_{\rm s}$, is included to correctly account for correlations with $R$, $l_{\rm A}$, and $\Omega_{\rm b0} h^2$.}

As discussed in \cite{Chen:2018dbv}, the CMB distance prior should be used to constrain models that deviate from the $\Lambda$CDM model at late times, and which are  expected to not significantly impact the peak structure of the CMB power spectrum. In our case, this corresponds to a negligible contribution from $\rho_{\rm comp}$ at early times, especially those that are proportional to $(1+z)^3$. Although this is \textit{a priori} guaranteed for our choice of priors, in Appendix~\ref{ap:consistency} we investigate whether the use of the CMB distance prior is consistent with the analysis. We address in particular the potential changes that $\rho_{\rm comp}$ induce in the definition of the shift parameter.

\subsection{SNIa}

In order to constrain late-time deviations from $\Lambda$CDM of the expansion rate when considering $\beta \in [0,10]$, we use the cosmological distance provided by standard candles. In particular, we use the Pantheon SNIa compilation \cite{Scolnic:2017caz}.

Standard candles measure the apparent magnitude $m$, which constrains the background dynamics of the Universe through the relation
\begin{equation}
    m(z) = 5 \log\frac{D_{\rm L}(z)}{1 \mathrm{Mpc}} + 25 + M \,,
\end{equation}
with $D_{\rm L}$ the luminosity distance, and $M$ the absolute magnitude of SNIa.

\begin{figure*}[!tbp]
    \centering
    \includegraphics[scale=0.8]{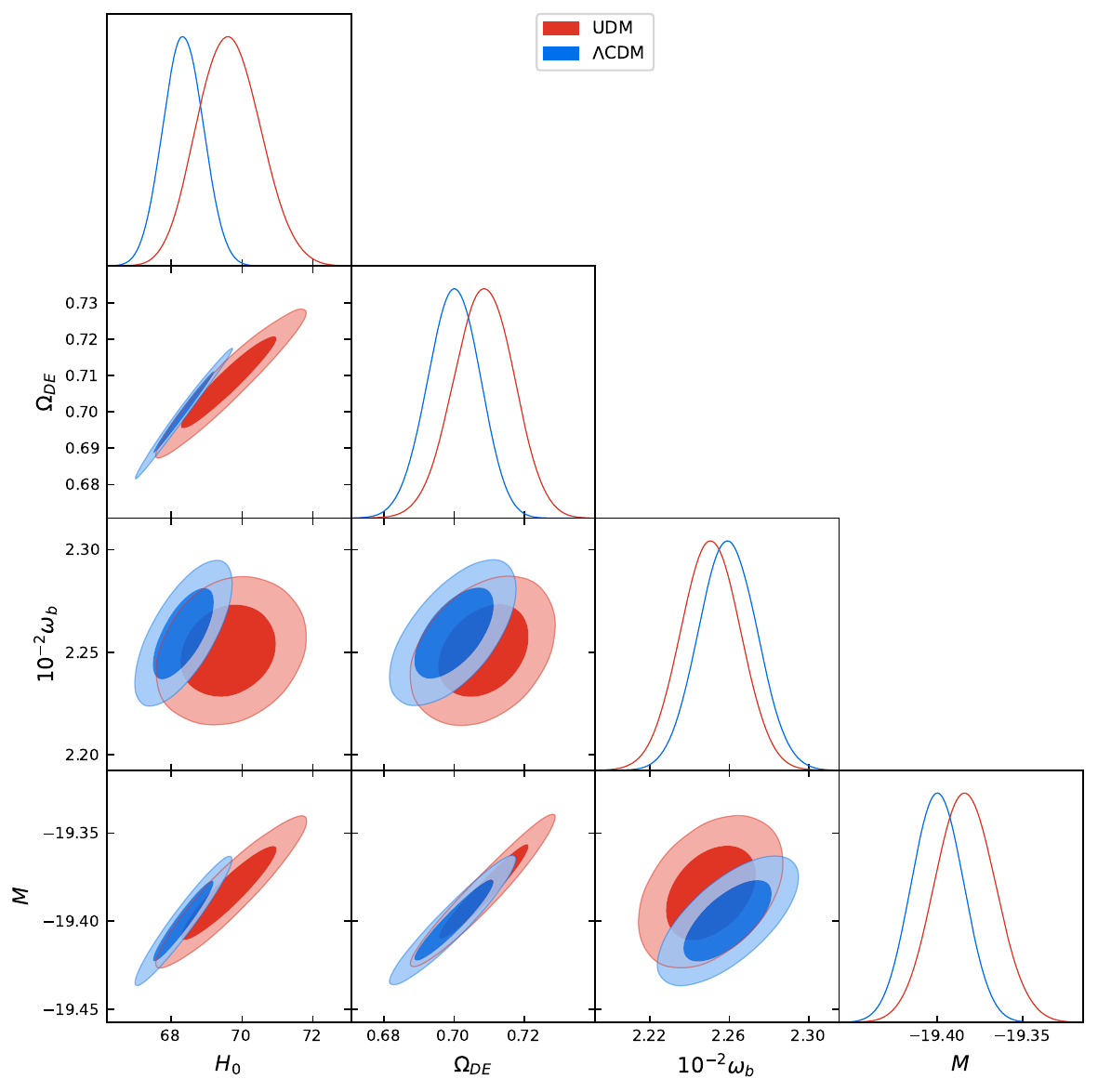}
    \caption{Marginalized constraints (68\% and 95\% credible regions) of the UDM and $\Lambda$CDM models from the Planck 2018 CMB prior, Pantheon supernovae and the local prior on the supernova absolute magnitude $M$.}
    \label{fig:udmvslcdm}
\end{figure*}

\subsection{Absolute Magnitude of SNIa}

In order to offer a calibration of the apparent magnitude of SNIa, we use the Gaussian prior on $M$
\begin{align}
    \chi^2 = \frac{(M-M_{\rm R21})^2}{\sigma_{M_{\rm R21}}^2} \;,
\end{align}
whose use is equivalent to the local determination of $H_0$ by SH0ES \cite{Riess:2020fzl}. Indeed, as discussed in \cite{Camarena:2021jlr}, the use of a prior on $M$ instead of a prior on $H_0$ provides several advantages. For instance, it counts for the discrepancy on the absolute magnitude measured by the CMB distances and the local Cepheids, and it also avoids the double counting of SNIa in the range $0.023 < z < 0.15$.

Lastly, it is important to note that our analysis does not include BAO data. Although standard rulers provided by BAO  strongly constrain late-time modifications of the $\Lambda$CDM, the interpretation of the clustering of matter and formation of BAO is incomplete without understanding the evolution of cosmological perturbations. On the other hand, as shown in the following section, the combination of the CMB priors and the SNIa already provide stringent constraints of the parameter space of the UDM model.

In order to understand the effect of the local determination of the Hubble function on the results, in the following, we perform the Bayesian analysis of the $\Lambda$CDM model and the UDM model considering two cases: one with and one without the prior on $M$. The comparison between the UDM and our $\Lambda$CDM analysis made in the same conditions will be the basis to assess the potential to alleviate the tension. We implement the background evolution of the UDM model in CLASS~\cite{Lesgourgues:2011re,Blas:2011rf}, and we perform the MCMC sampling with MontePython~\cite{Audren:2012wb,Brinckmann:2018cvx}. We produce most of the plots of this section using GetDist~\cite{Lewis:2019xzd}. The modified CLASS version can be accessed at \href{https://github.com/EFrion/class_public}{github.com/EFrion/class\_public}. The MCMC analysis  uses the cosmological parameters $\left\lbrace \omega_{\rm b0},~\omega_{\rm cdm0},~ n_{\rm s},~h,~M\right\rbrace$ with improper flat prior, plus the two parameters $\left\lbrace \beta, a_{\rm t} \right\rbrace$ of the UDM model, whose flat priors are shown in Table~\ref{table_params}. To assess the convergence of model parameters, we assume that the Gelman--Rubin convergence criterion \cite{Gelman:1992zz} is $R-1<10^{-3}$ for each parameter.

\begingroup 

\setlength{\tabcolsep}{10pt} 
\renewcommand{\arraystretch}{1.5} 
\begin{table}

\begin{tabularx}{\linewidth}{ X X }

Parameter & Prior \\
\hline \hline

{\boldmath$\beta{}_{udm } $} & $\left[ 10^{-3}, 10 \right]$ \\

{\boldmath$a_{\rm t }$} & $\left[ 0, 1 \right]$ \\

\hline \hline

\end{tabularx}

\caption{UDM parameters used in the MCMC simulation.}

\label{table_params}
\end{table}

\endgroup

\begin{figure*}[!tbp]
    \centering
    \includegraphics[scale=0.8]{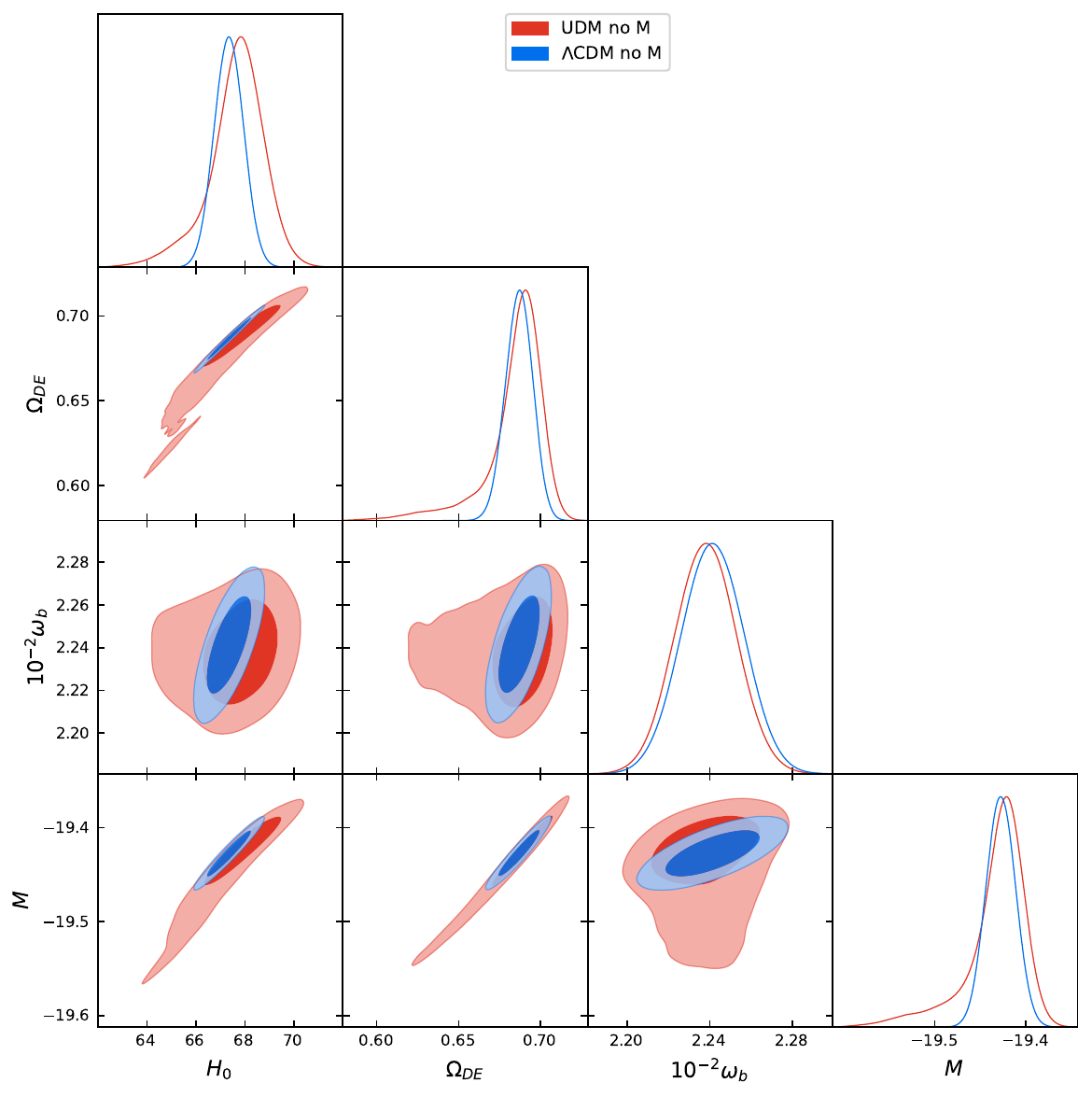}
    \caption{Marginalized constraints (68\% and 95\% credible regions) of the UDM and $\Lambda$CDM models from the Planck 2018 CMB prior and Pantheon supernovae.}
    \label{fig:udmvslcdm_noM}
\end{figure*}

\begin{figure*}[!tbp]
    \centering
    \includegraphics[scale=0.7]{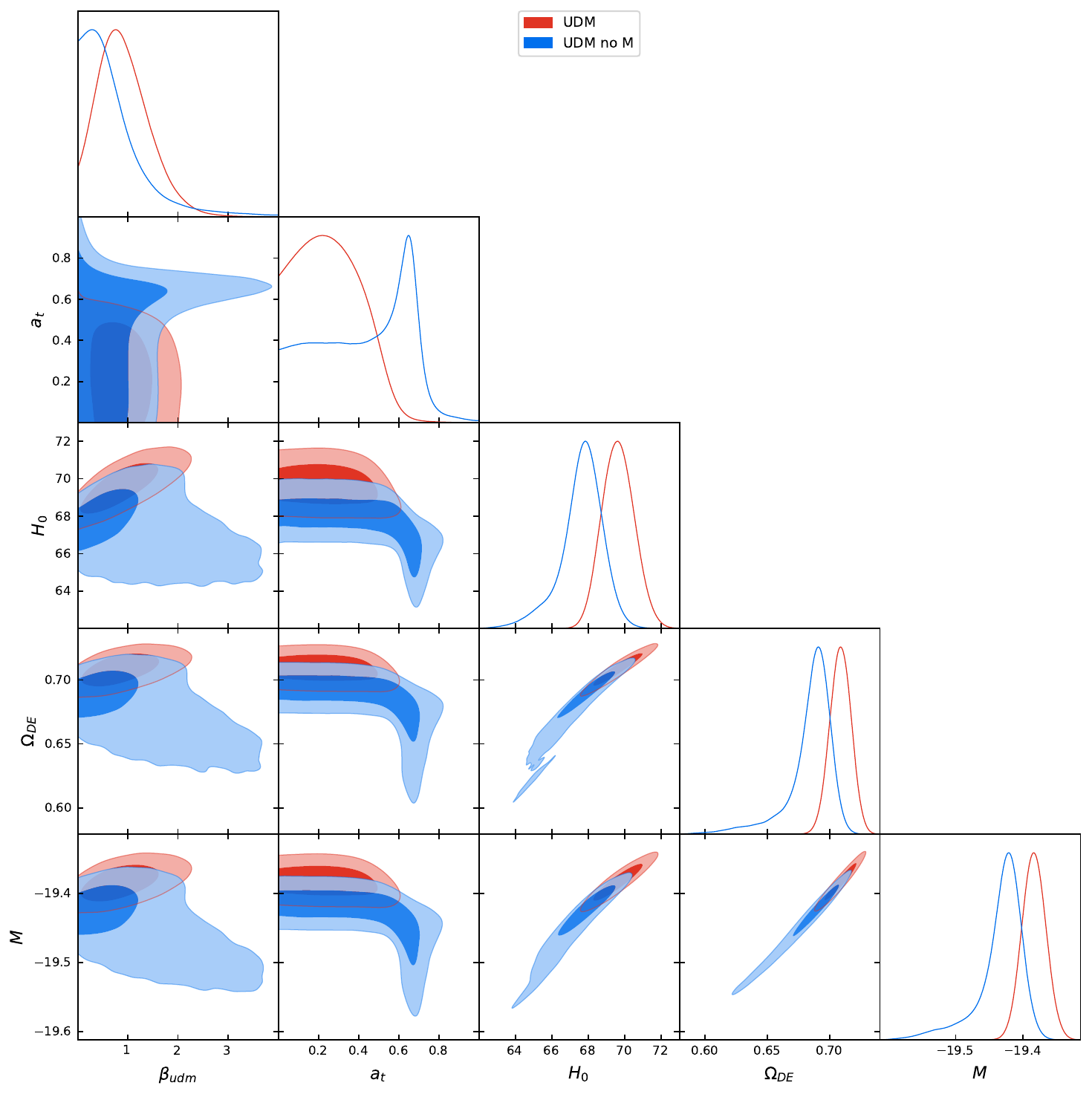}
    \caption{Marginalized constraints (68\% and 95\% credible regions) of the UDM model from the Planck 2018 CMB prior and Pantheon supernovae, either with and without the local prior on the supernova magnitude $M$.}
    \label{fig:udmvsudm_noM}
\end{figure*}

\begin{figure}[tbp]
    \centering
    \includegraphics[width=\columnwidth]{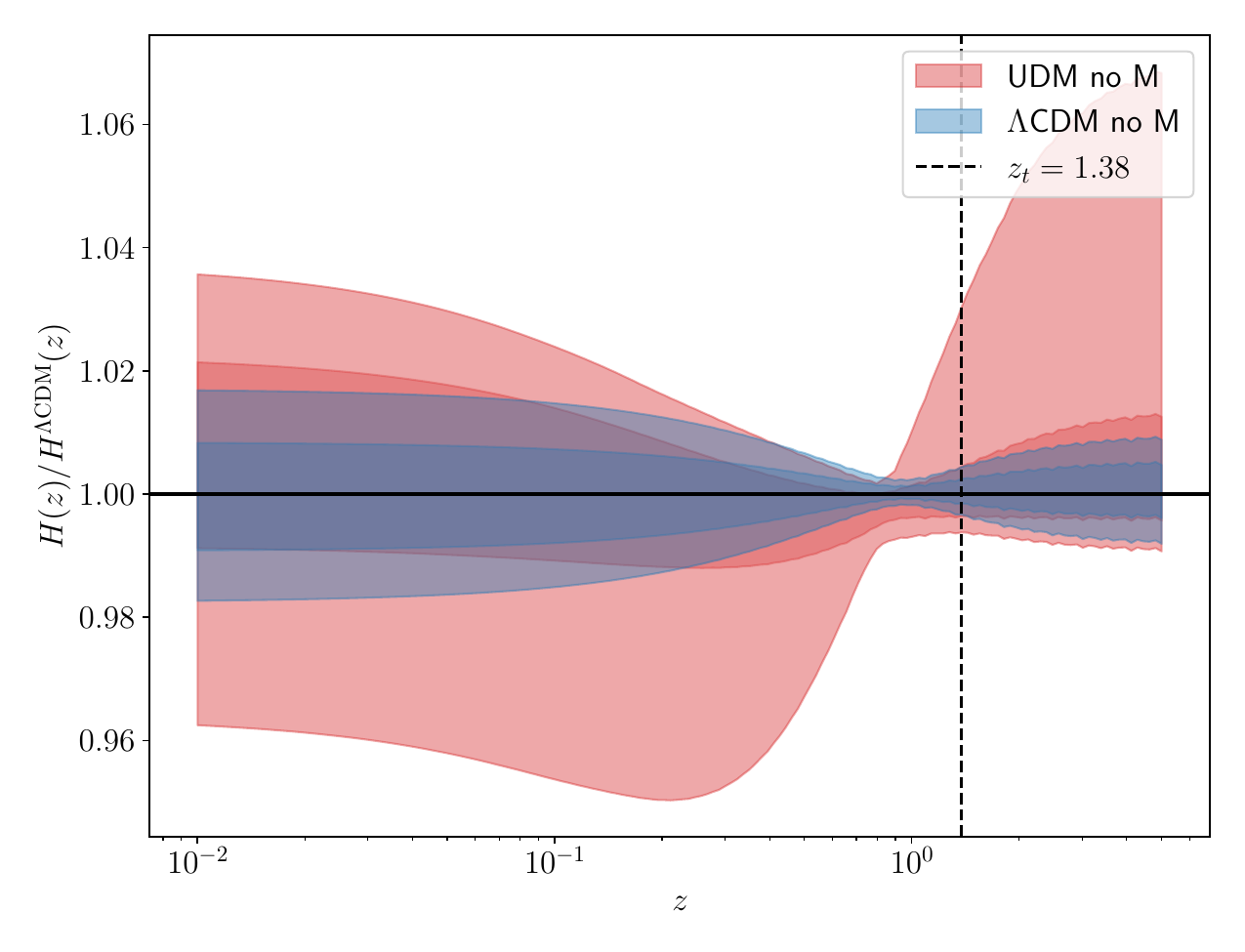}
    \includegraphics[width=\columnwidth]{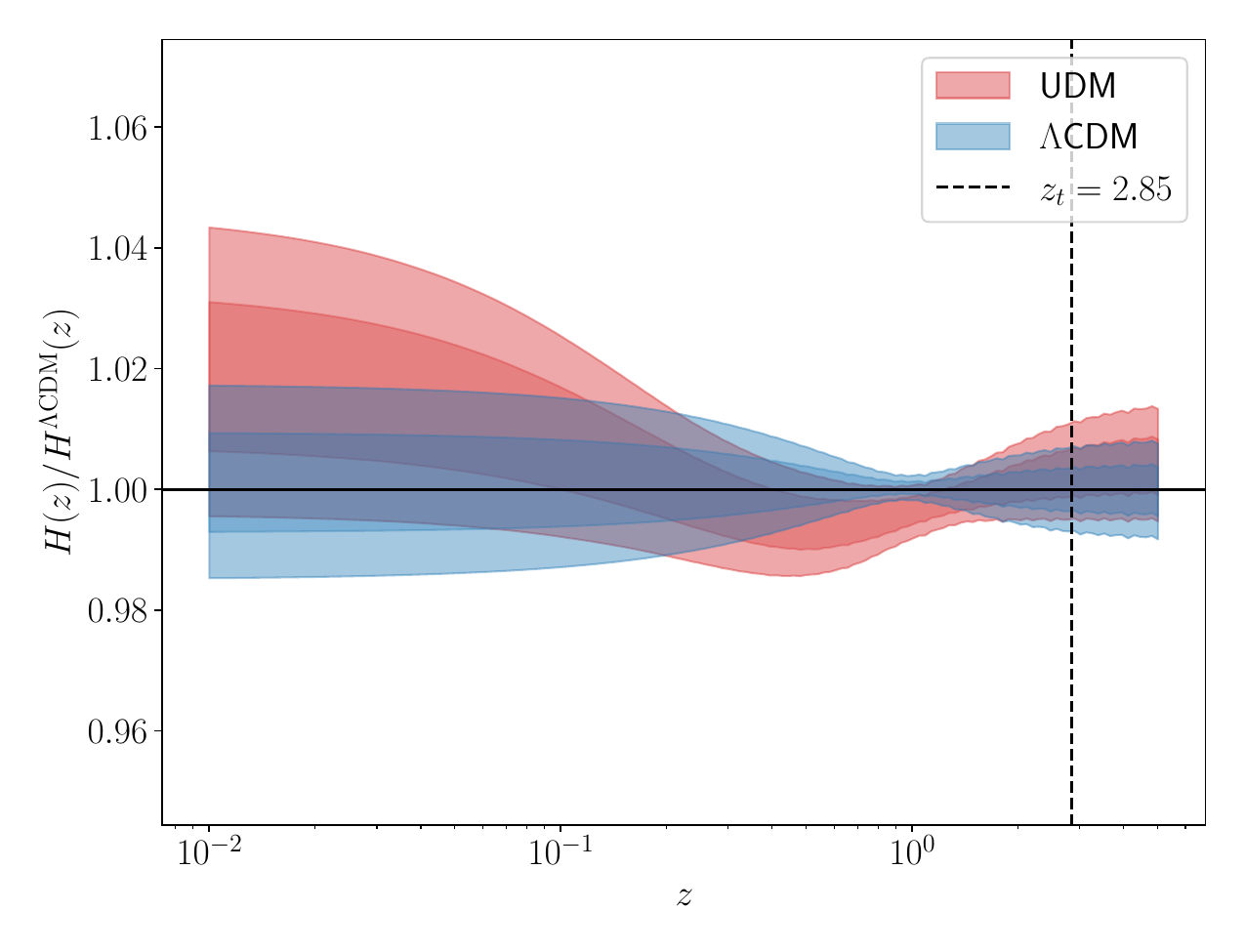}
    \caption{Ratio between the Hubble function obtained through cosmological constraints and the expansion rate predicted by the $\Lambda$CDM baseline, from Table~\ref{table_marg}. The bottom panel includes the prior on $M$, the top panel does not. In both panels, the vertical dashed line indicates the preferred redshift transition, which is slightly higher with the prior on $M$, $z_t=2.88$ vs $z_t=1.36$ without it.}
    \label{fig:ratio_Hz_constrains}
\end{figure}

\section{Results}\label{results}

\begingroup 

\setlength{\tabcolsep}{20pt} 
\renewcommand{\arraystretch}{1.7} 
\begin{table*}

\begin{tabularx}{\textwidth}{lcccc}

 Parameter & UDM (68\%) & $\Lambda$CDM (68\%) & UDM no $M$ (68\%) & $\Lambda$CDM no $M$ (68\%) \\
\hline \hline
{\boldmath$\beta{}_{\rm udm } $} & $0.93^{+0.38}_{-0.62}$ & - & $<0.862 $ & -\\

{\boldmath$a_{\rm t }$} & $0.26^{+0.12}_{-0.21}$ & - & $0.42\pm 0.22 $ & - \\

{\boldmath$10^{-2}\omega{}_{\rm b }$} & $2.251\pm 0.015   $ & $2.259\pm 0.015   $ & $2.238\pm 0.015   $ & $2.241\pm 0.015   $\\

{\boldmath$M $ [mag] } & $-19.384\pm 0.018 $ & $-19.400\pm 0.015  $ & $-19.434^{+0.037}_{-0.014} $ & $-19.427\pm 0.016 $ \\

{\boldmath$H_0 $} [km/s/Mpc] & $69.64\pm 0.88        $ & $68.34\pm 0.57       $ & $67.6^{+1.3}_{-0.82}        $ & $67.35\pm 0.60       $ \\

{\boldmath$\Omega_{\rm DE0} $} & $0.7084\pm 0.0085   $ & $0.6999\pm 0.0074   $ & $0.684^{+0.019}_{-0.0067}   $& $0.6868\pm 0.0083   $ \\

\hline\hline

\end{tabularx}

\caption{Marginalized constraints of the UDM and $\Lambda$CDM models.}

\label{table_marg}
\end{table*}

\endgroup

\begingroup 

\setlength{\tabcolsep}{30pt} 
\renewcommand{\arraystretch}{1.7} 
\begin{table*}[!t]

\begin{tabularx}{2\columnwidth}{lcccc}

 Parameter & UDM & $\Lambda$CDM & UDM no $M$  & $\Lambda$CDM no $M$  \\
\hline \hline
{\boldmath$\beta{}_{\rm udm } $} & $0.80$ & - & $0.339 $ & - \\

{\boldmath$a_{\rm t }$} & $0.14$ & - & $0.417 $ & - \\

{\boldmath$10^{-2}\omega{}_{\rm b }$} & $2.251   $ & $2.258   $ & $2.239   $ & $2.242 $\\

{\boldmath$M $ [mag] } & $-19.383$ & $-19.40 $ & $-19.422 $ & $-19.427 $ \\

{\boldmath$H_0 $} [km/s/Mpc] & $69.60 $ & $68.26       $ & $67.82        $ & $67.38       $ \\

{\boldmath$\Omega_{\rm DE0} $} & $0.7088 $ & $0.6990$ & $0.6900   $& $0.6873   $ \\

\hline \hline

{\boldmath$\chi_{\rm Pant}^2$} & $1028.2$ & $1025.9$ & $1025.6$ & $1026.0$ \\
{\boldmath$\chi_{\rm cmb}^2$} & $9.9 \times 10^{-1}$ & $2.9$ & $3.9 \times 10^{-2}$ & $9.5 \times 10^{-2}$ \\
{\boldmath$\chi_{\rm M}^2$} & $14.5$ & $18.0$ &  &  \\
{\boldmath$\chi_{\rm tot}^2$} & $1043.6$ & $1046.8$ & $1025.7$ & $1026.1$ \\

\hline\hline

\end{tabularx}

\caption{Best fit of the UDM and $\Lambda$CDM models. We display below their $\chi^2$ for each individual experiment and their combined sum.}

\label{tablebf}
\end{table*}

\endgroup

Given our assumptions, we are in a position to constrain unambiguously the UDM model. In the presentation of these results, we denote the density of the effective dark energy component today as $\Omega_{\rm DE0}$, whether it comes from vacuum (in the $\Lambda$CDM case) or from the component $\rho_{\mathrm{\rm DE}} = \rho_\varphi - \rho_{{\rm m}0}a^{-3}$ (in the UDM case). Introducing the prior on $M$ effectively mixes early probes (CMB) with late-time probes (SNIa), which may \textit{a priori} be in tension. We check in Figure \ref{fig:udmvsudm_noM} that removing the prior leads to the same conclusions.

\textbf{UDM vs $\Lambda$CDM} -- In figure \ref{fig:udmvslcdm}, we present the constraints for the set of variables $\left\lbrace H_0,~\Omega_{\rm DE0},~\Omega_{\rm b0},~M\right\rbrace$ with a prior on the absolute magnitude $M$. In the UDM scenario, the dark energy contribution from the complementary fluid is typically greater than the vacuum energy density in $\Lambda$CDM which reflects in a slightly greater value for $H_0$.  From Table~\ref{table_marg}, we can read the constraint $H_0=69.64\pm 0.88$ in the UDM case, which is indeed  bigger than in the $\Lambda$CDM case, for which it is $H_0=68.3^{+1.1}_{-1.1} $, though the difference is modest.

\textbf{UDM vs $\Lambda$CDM (no $M$)} --
Fig.~\ref{fig:udmvslcdm_noM} shows that, in the absence of the prior on $M$, the constraints on the UDM scenario loosen up while the constraints on $\Lambda$CDM are still tight. Since the SNIa are effectively calibrated by the CMB distance, the UDM model reproduces a cosmic expansion consistent with $\Lambda$CDM and the best fit values of the UDM model parameters are very close to those of  $\Lambda$CDM. Additionally, Table~\ref{table_marg} shows that constraints on the UDM scenario leads to almost negligible increase on $H_0$ together with a significant increase in the uncertainties.  
This exemplifies that the $M$ prior helps in constraining more precisely the dark energy content in alternative scenarios to the standard cosmological model.

\textbf{UDM vs UDM no $M$} -- The effect of the prior on $M$ is even more striking when comparing the Bayesian analysis for the UDM model with and without the prior on $M$. Figure \ref{fig:udmvsudm_noM} gives a clear visual proof that the prior helps to constrain the UDM model, and in particular the two parameters $\beta$ and $a_t$ specific to the EoS transition. We see that the prior enhances the best fit values of $\beta$, $H_0$, $\Omega_{\rm DE0}$, and $M$, while decreasing $a_t$. In both cases, the value for $\beta$ is quite small ($<2$ in the $1\sigma$ region), in favour of a smooth transition. The prior favours an earlier transition redshift, which explains the highest best fit for $H_0$ ($69.64 \pm 0.88$ against $67.6^{+1.3}_{-0.82}$ without the prior). 

The tension between the $\Lambda$CDM analyses presented here and the observed value of $H_0$~\cite{Riess:2021jrx} is about $\sim 4\sigma$, while the tension between the UDM model and $H_0$ is about $\sim 2.5\sigma$. Therefore, the UDM model reduces the tension by $\sim 1.5\sigma$. The tension is reduced whether we include the prior on $M$ or not.

Table~\ref{tablebf} gives the best-fit values together with the $\chi^2$ for each individual experiment and their combined sum. The relative difference in the Hubble rate with respect to $\Lambda$CDM and the equation of state of the dark energy-like component are shown in Appendix~\ref{ap:bf}.

In Fig.~\ref{fig:ratio_Hz_constrains}, we compare the evolution of the Hubble rate extracted from the Bayesian analysis for the UDM and $\Lambda$CDM models using Table~\ref{table_marg}. The bottom panel shows that, when $M$ is included in the analysis, only a slight deviation from the $\Lambda$CDM regime is allowed by data inducing to a $2\%$ increase in the Hubble constant. On the other hand, as shown by the top panel, when $M$ is not assumed in the analysis, the UDM model displays a cosmic expansion consistent with the $\Lambda$CDM model and even smaller deviation are allowed by data, although, as expected,  uncertainties are larger. We find that the preferred redshift transition is slightly higher with the prior on $M$, $z_t=2.88$ vs $z_t=1.36$ without it.

\begin{figure*}[!tbp]
\centering
	\includegraphics[width=\textwidth]{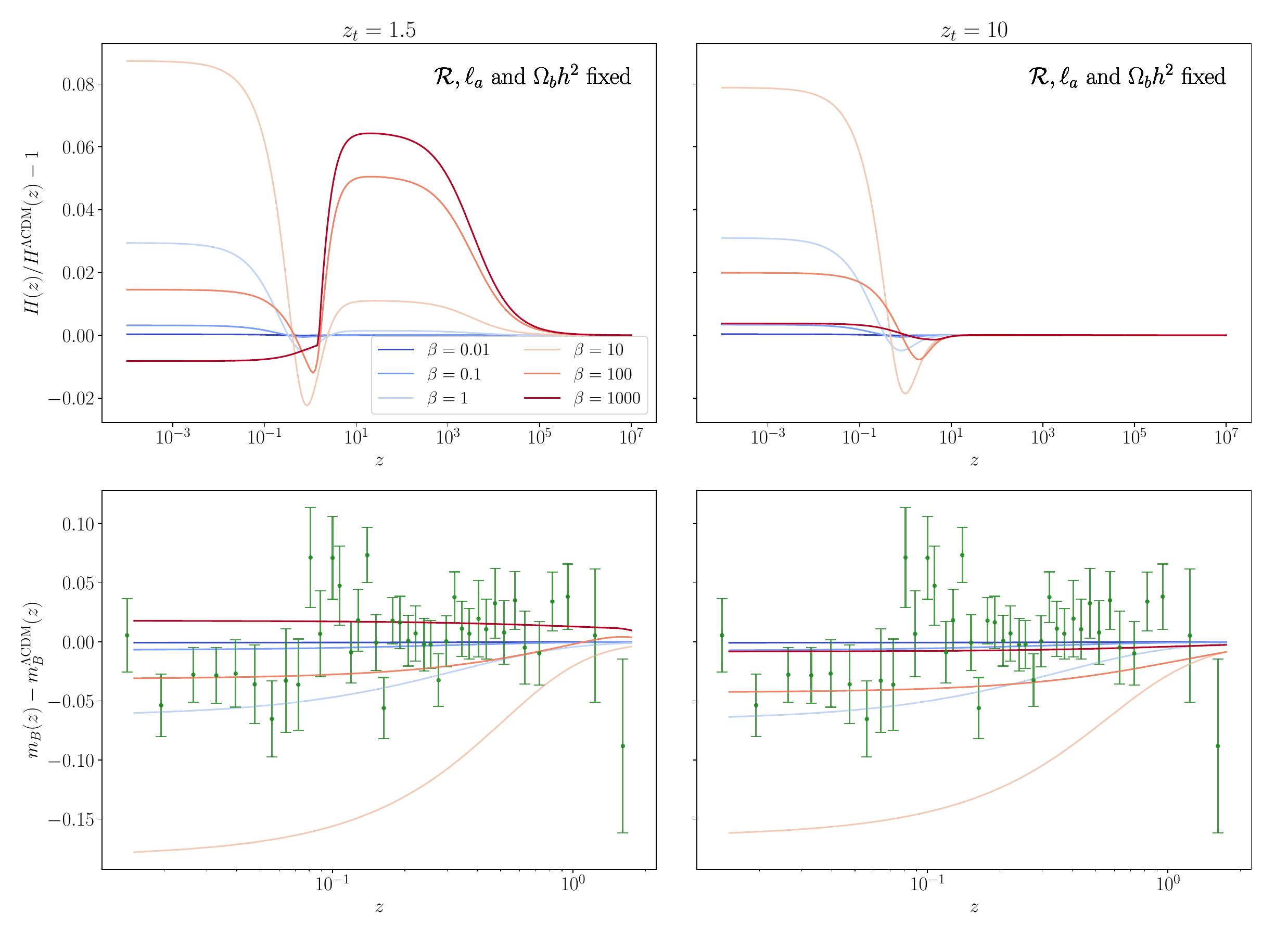}\label{fig:1.1}\\
	\caption{Top panels: relative variation of the Hubble rate when the quantities related to the CMB priors are assumed to be fixed. Bottom panels: residuals for the apparent magnitude between the theoretical prediction of the $\Lambda$CDM model and Pantheon observations in its binned version (green dots). We also display the difference between the $\Lambda$CDM and UDM models for different values of $\beta$. The UDM model for $\beta=10$, the most promising case for solving the tension, is inconsistent with observations.}
	\label{fig:1}
\end{figure*}

In Fig.~\ref{fig:1}, the top panels show how the Hubble rate varies when we keep fixed the CMB priors $R$ and $l_{\rm A}$, as well as the physical baryon density. The panels are very similar to the theoretical prediction from Fig.~\ref{fig:DH_dA_fix}, and confirm that $\beta \in [1,10]$ is required to alleviate the tension. The Bayesian analysis in Figs.~\ref{fig:udmvslcdm}, \ref{fig:udmvslcdm_noM}, and \ref{fig:udmvsudm_noM}  provide the stronger constraints $\beta<0.93^{+0.38}_{-0.62}$ with the $M$ prior and $\beta<0.862$ without, which is explained when considering the expected difference in supernovae magnitude. The bottom panels show this for the two models. The green dots with error bars are binned Pantheon measurements. In these panels, the UDM model with $\beta \approx 10$ (pink line) is inconsistent with observations, while the slow and fast transition regimes are consistent. 

We complement the previous results with two model selection criteria, namely the Akaike information criterion (AIC) \cite{akaike1974new} and the Bayesian information criterion (BIC) \cite{schwarz1978estimating}. They are defined as
\begin{align}
    \textup{AIC} &= \chi^2_{\rm min} + 2k \;, \\
    \textup{BIC} &= \chi^2_{\rm min} + k \ln{N} \;,
\end{align}
where $k$ is the number of parameters of a model and $N$ is the number of data points used to derive the probabilities of the parameters. Summing the data from the Pantheon catalog (1048) and the CMB priors (3) results in  $\ln{N}=7$, regardless of the prior on $M$. We compare the UDM and $\Lambda$CDM models with the differences \cite{Camarena:2018nbr}
\begin{align}
    \Delta AIC &= \Delta \chi^2 + 2 \Delta k \;, \\
    \Delta BIC &= \Delta \chi^2 + \Delta k \ln{N} \;,
\end{align}
in which a positive value means the $\Lambda$CDM model is favoured over the UDM model. The results reported in Table~\ref{criteria} are positive, both in the case assuming a prior on $M$ or not. Therefore, we conclude that the $\Lambda$CDM is favoured. According to the qualitative interpretations of the  criteria found in Table VI and VII of Ref~\cite{Camarena:2018nbr}, the empirical support of $\Lambda$CDM is substantial ($\Delta$AIC$<2$), and the evidence against the UDM model is very strong ($\Delta$BIC$>10$).

\begingroup 

\setlength{\tabcolsep}{10pt} 
\renewcommand{\arraystretch}{1.5} 
\begin{table}
\centering
\begin{tabular}{c c c}

 Criterion &  With prior & Without prior \\

\hline \hline

{\boldmath$\Delta AIC$} & $0.8$ &  $3.6$  \\
{\boldmath$\Delta BIC$} & $10.8$  & $13.6$  \\

\hline

\end{tabular}

\caption{Difference of the Akaike (AIC) and Bayesian (BIC) information criteria between the UDM and $\Lambda$CDM models. The selection criteria favour the $\Lambda$CDM model, regardless of the prior on $M$.}
\label{criteria}
\end{table}

\endgroup

\section{Discussion}
\label{sec:concl}

We performed in this work an analysis of a UDM model that acts also as dark energy at late times.  Two goals were achieved: 1)~we analysed the pressure profile~\eqref{pUDM} chosen in \cite{Bertacca:2010mt} with a Bayesian approach using CLASS and MontePython for the first time, and 2)~we assessed whether this profile possesses the capability of alleviating the Hubble tension. This particular profile is similar to a late-time modification of the $\Lambda$CDM model in which the transition from the matter regime to the dark energy regime is parametrised by two variables, $\beta$ and  $a_t$. In Section \ref{sec:udm}, we argued that the product $\beta (a^3-a_t^3)$ controls the behaviour of the transition. If the product is very small, the transition happens smoothly and, conversely, it happens quickly if the product is large.

We find that supernovae data constrain $\beta$ in the range $0<\beta<2$, though $a_t$ is  less constrained. The correlation of $a_t$ with independent variables such as $H_0$, the dark energy density today $\Omega_{\rm DE0}$ or the supernovae absolute magnitude $M$ gives $a_t<0.42\pm 0.22 $ at $1\sigma$, or  $a_t<0.26^{+0.12}_{-0.21}$ when a prior on $M$ is assumed. This implies that the transition has to occur at least at a redshift $z>0.5$. The posteriors on $a_t$ hint at a transition redshift $z_t=1.38$ when no prior is assumed, and a further transition in time at $z_t=2.85$ with the prior. In light of the preferred values of $\beta$ and $a_t$, we conclude that the data favours a smooth transition ($\beta (a^3-a_t^3)<1$) over a quick one. 

Although our analyses demonstrate that the UDM model can accommodate a higher value of $H_0$, the displacement observed with respect to the $\Lambda$CDM model is subdominant. The main factor leading to $\sim 1.5\sigma$ decrease of the Hubble tension is an increase of  uncertainties associated with correlated parameters. Additionally, the AIC (BIC) information criterion penalizes the UDM model and points out to a substantial (strong) empirical support to the $\Lambda$CDM model. Thus, overall, even in the most promising case, \textit{i.e., assuming a prior on $M$}, the UDM model does not constitute an advantageous solution to the Hubble tension.

Our results partially agree with the claims made in Ref~\cite{Lee:2022cyh}, in which the authors argue that a DE model must possess two features in order to potentially solve the Hubble tension. First, the equation of state must cross the phantom line $w_{\rm DE}< -1$, and, second, the integrated dark energy density must be smaller than that of a cosmological constant in $\Lambda$CDM.
As we discuss in Appendix~\ref{ap:intde}, the UDM model possesses both these requirements, however both lower and higher values of $H_0$ are allowed, where the typical increase in $H_0$ is not enough to explain away the tension. On the other hand, this does not necessarily mean that this UDM model is unable to solve the tension. Indeed, we only consider the background evolution of the model in this analysis, even though perturbations can also affect the equation of state of the complementary component and potentially have an impact on $H_0$. We will investigate this possibility in a future paper.

\section*{Acknowledgments}

EF thanks the Helsinki Institute of Physics for their hospitality. The numerical analysis was done using the Puck cluster of the University of Jyväskylä. DC thanks the Robert E.~Young Origins of the Universe Chair fund for its generous support. DB acknowledges support from the COSMOS network (www.cosmosnet.it) through the ASI (Italian Space Agency) Grants 2016-24-H.0, 2016-24-H.1-2018 and 2020-9-HH.0.
VM thanks CNPq (Brazil, 307969/2022-3) and FAPES (Brazil, TO 365/2022, 712/2022, 976/2022, 1020/2022, 1081/2022) for partial financial support. LG acknowledges support from the Australian Government through the Australian Research Council Laureate Fellowship grant FL180100168.

\bibliographystyle{apsrev4-1}
\bibliography{main}

\twocolumngrid

\begin{appendix}

\section{Best fits }
\label{ap:bf}

\begin{figure}
    \includegraphics[width=\columnwidth, height=10cm]{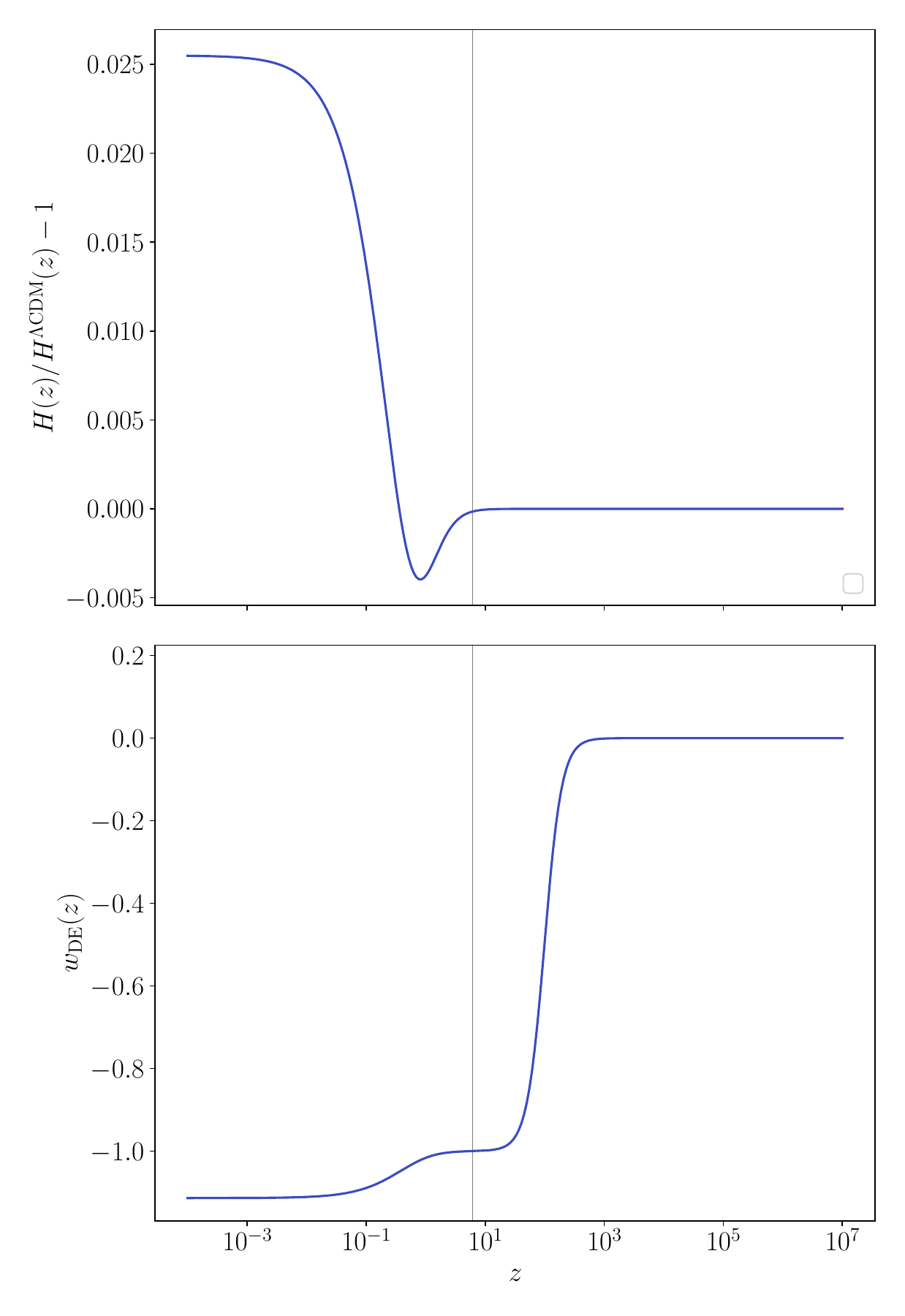}
    \caption{
    Relative difference in the Hubble rate wrt $\Lambda$CDM and the equation of state of the dark energy-like component using the best fit of the UDM model with prior on $M$ from Table~\ref{tablebf}.}
    \label{fig:bestfits}
\end{figure}

We display in this section the difference in the Hubble rate and the equation of state of the dark energy-like component using the best fit of the UDM model with prior on $M$, \textit{i.e.}, the first column of Table~\ref{tablebf}. They are shown in Figure \ref{fig:bestfits}.

\section{Integrated dark energy}
\label{ap:intde}

\begin{figure}[t]
    \centering
    \includegraphics[width=\columnwidth]{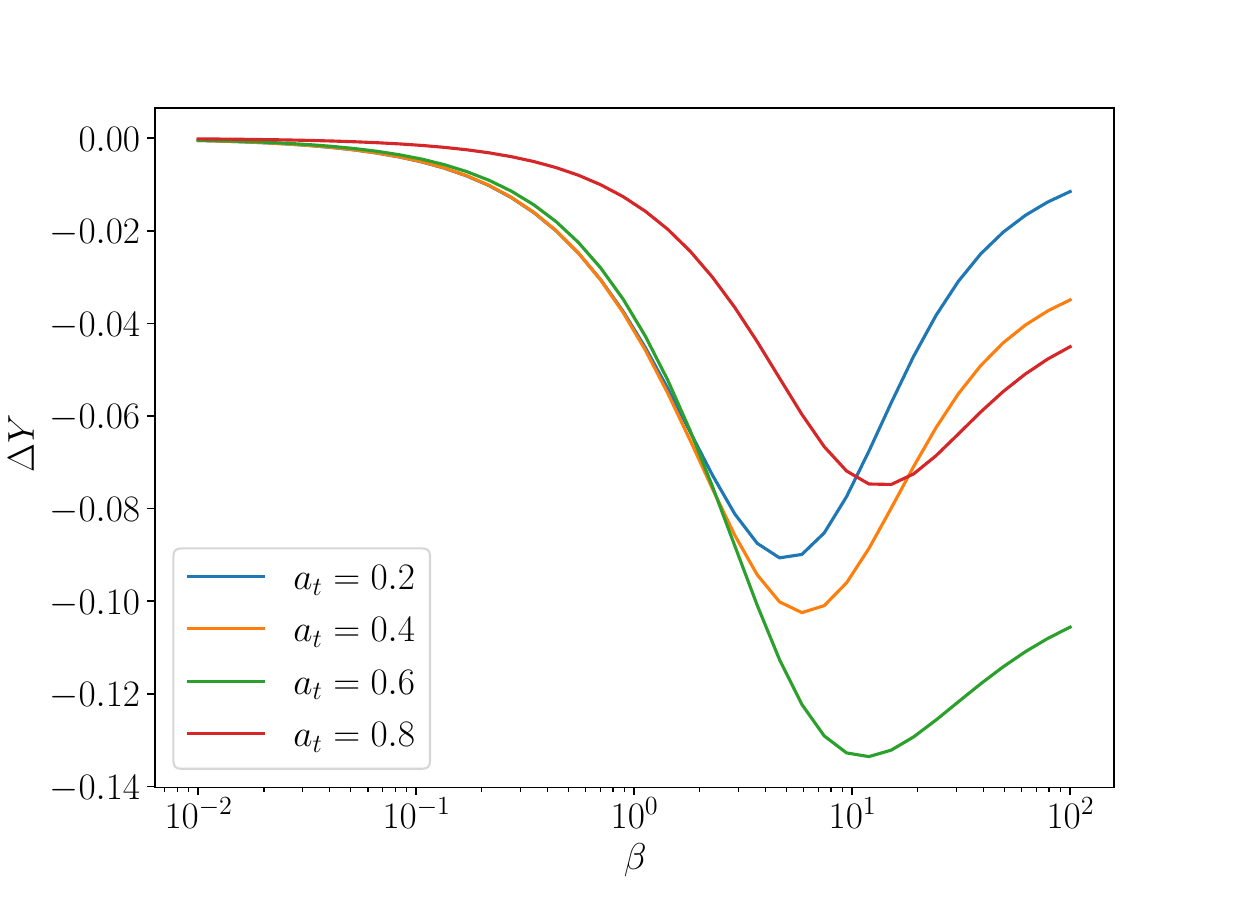}
    \caption{Integrated dark energy difference for the UDM model.}
    \label{fig:deltay}
\end{figure}

\cite{Lee:2022cyh} argues that a DE model must possess two features in order to potentially solve the Hubble tension. First, the equation of state must cross the phantom line $w_{\rm DE}< -1$, and, second, the integrated dark energy density must be smaller than that of a cosmological constant in $\Lambda$CDM.
We have already seen in Fig.~\ref{fig:wz_wec} that the UDM model possesses an effective phantomic dark energy component at late times.
Regarding the integrated dark energy density, we show in Figure \ref{fig:deltay} the quantity  $\Delta Y (z_{\star})$, which is the integrated dark energy contribution with respect to $\Lambda$CDM, where $z_{\star}$ is the redshift at which dark energy transitions from a decelerated phase to an accelerated one or vice-versa~\cite{Lee:2022cyh}. For our model, we find that $z_{\star} \simeq 0.64$, which is approximately the expected value that should be consistent with Planck data for $\Lambda$CDM. We display the integrated dark energy for our UDM model for $\beta \in \left[10^{-2},10^2\right]$ and $a_{\rm t} \in \left\lbrace 0.2, 0.4, 0.6, 0.8 \right\rbrace$, in which we show it is always negative.
Although this Figure shows that most of the parameter space of the UDM model generates a negative $\Delta Y$, it does not display the correlation between $\Delta Y$ and $H_0$ found in \cite{Lee:2022cyh}. The reason is that \cite{Lee:2022cyh} considered CPL models which feature a qualitatively different phenomenology as compared to the UDM model considered in this study.

\begin{figure}[t]
    \centering
    \includegraphics[width=\columnwidth]{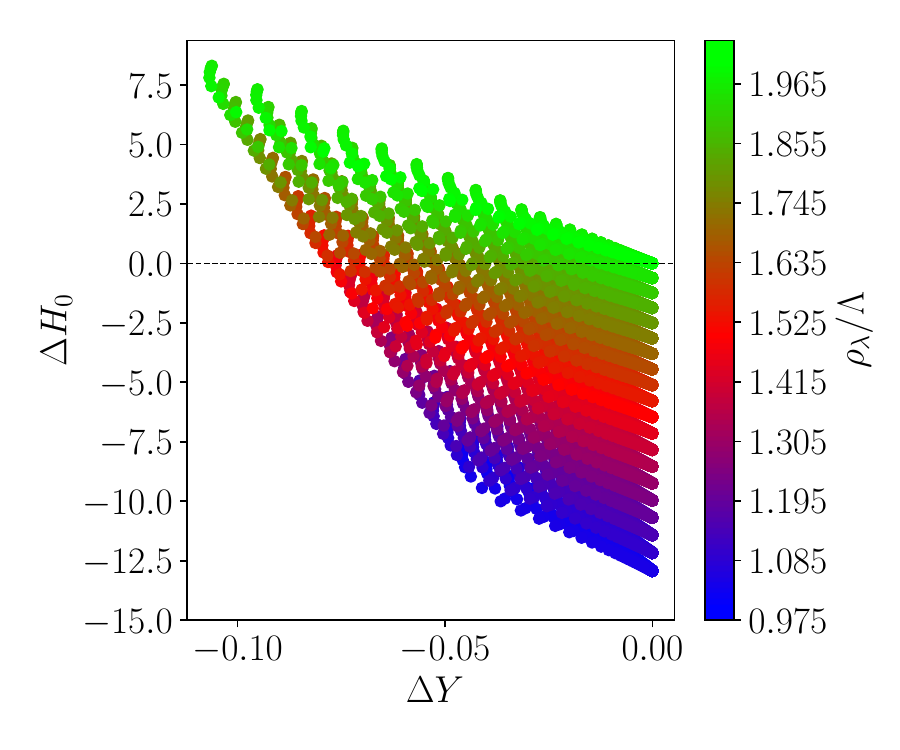}
    \caption{Correlation between the integrated dark energy difference and the Hubble constant considering 18000 realizations of the UDM model in the region $\rho_\lambda \in [\Lambda, 2\Lambda]$, $a_t \in [0.075,0.65]$ and $\beta \in [0,2.5]$.}
    \label{fig:deltay_dh0}
\end{figure}

To unveil the possible correlation between the integrated dark energy and the Hubble constant, we compute $\Delta Y$ and $\Delta H_0$ for 18000 different combinations of the UDM parameters. Specifically, we use points of the UDM parameter space that approximately reassemble $95\%$ credible region of Fig.~\ref{fig:udmvsudm_noM}, this is, we adopted points in the region $a_t \in [0.075,0.65]$ and $\beta \in [0,2.5]$. Furthermore, we consider values of the energy density of the effective cosmological
constant in the range $\rho_\lambda \in [\Lambda, 2\Lambda]$. The results of this sampling are shown in Fig.~\ref{fig:deltay_dh0}.

In concordance with Fig.~\ref{fig:deltay}, Fig.~\ref{fig:deltay_dh0} shows that most of the UDM parameter space is consistent with $\Delta Y < 0$. Nevertheless, the latter does not ensure an increase in the Hubble constant. Indeed, given the strong correlation with $\rho_\lambda$, $\Delta H_0$ spans over positive and negative values showing that most of the parameter space allowed by the data does not lead to a higher value of the Hubble constant. It is important to stress that the constraints displayed in Table~\ref{table_marg} imply a mean value of $\rho_\lambda \approx 1.5 \Lambda$ (reddish points in Fig.~\ref{fig:deltay_dh0}), for which $\Delta H_0$ is mostly negative.

\section{Consistency analysis: CMB priors}
\label{ap:consistency}

When implementing the model in CLASS, it is  necessary to understand whether the UDM energy density \eqref{rhophi} behaves as ordinary matter at recombination. This is crucial in order to define the fluid initial conditions. Since the log term of \eqref{rhophi} possesses a complex behaviour, we must know if it contributes to the usual matter term decaying as $a^{-3}$. In the eventuality that it does, the shift factor $R(z)$ \eqref{shift_factor} would be affected, modifying one of the CMB priors. To ensure this is not the case, we must make sure the contributions from the matter energy density, $\rho_{\rm m0} a^{-3}$, and the complementary fluid, $\frac{3\rho_{\rm \lambda}}{2\beta} \log \left\lbrace \cosh{\left[\frac{\beta}{3} (a^3-a_t^3\right]}\right\rbrace a^{-3}$, are negligible at recombination. 

Let us call $\Omega_{\rm m}^{rec}$ and $\Omega_{\rm comp}^{rec}$ their respective energy densities, and $\delta \Omega := \Omega_{\rm comp}^{rec}/\Omega_{\rm m}^{rec}$ their ratio. Thus, when accounting for the eventual contribution of the complementary component, the difference in the shift factor \eqref{shift_factor} is equal to $\sqrt{1+\delta \Omega}$ and the percentage error produced in $R$ follows
\begin{align}
    \Delta R := 100 \big{|}1 - \sqrt{1+\delta \Omega} \big{|} \;.
\end{align}
Using the constraints derived from the analysis, we compute $\Delta R$. We find that all combinations of data lead to $\Delta R$ at least one order of magnitude smaller than the error of the CMB prior, \textit{i.e.}, the parameter space allowed by the data leads to $\Delta R \sim 0.02 \%$. This is consistent with the priors imposed in Section~\ref{sec:stat} and the assumption of exploring the UDM model as a late-time modification of the $\Lambda$CDM dynamics.

\end{appendix}

\end{document}